\documentclass[aps,showpacs,superscriptaddress]{revtex4}

\usepackage{times}
\usepackage{euscript}
\usepackage{amsfonts}
\usepackage{amsmath}
\usepackage{amssymb}
\usepackage{color}
\usepackage{graphicx}
\usepackage{epsfig}

\newcommand{\erf}{\mathop{\mathrm{erf}}\nolimits}
\newcommand{\Rea}{\mathop{\mathrm{Re}}\nolimits}
\newcommand{\Ei}{\mathop{\mathrm{Ei}}\nolimits}

\begin{document}

\title{Cooling force on ions in a magnetized electron plasma}

\author{Hrachya B. Nersisyan}
\email{hrachya@irphe.am}
\affiliation{Plasma Theory Group, Institute of Radiophysics and Electronics, 0203 Ashtarak, Armenia}

\author{G\"{u}nter Zwicknagel}
\email{guenter.zwicknagel@physik.uni-erlangen.de}
\affiliation{Institut f\"ur Theoretische Physik II, Universit\"{a}t
Erlangen-N\"{u}rnberg, Staudtstra{\ss}e 7, D-91058 Erlangen, Germany}
\date{\today }

\begin{abstract}
Electron cooling is a well-established method to improve the phase space quality of ion
beams in storage rings. In the common rest frame of the ion and the electron beam the ion
is subjected to a drag force and it experiences a loss or a gain of energy which eventually
reduces the energy spread of the ion beam. A calculation of this process is complicated as
the electron velocity distribution is anisotropic and the cooling process takes place in
a magnetic field which guides the electrons. In this paper the cooling force is calculated
in a model of binary collisions (BC) between ions and magnetized electrons, in which the
Coulomb interaction is treated up to second-order as a perturbation to the helical motion
of the electrons. The calculations are done with the help of an improved BC theory which
is uniformly valid for any strength of the magnetic field and where the second-order two-body
forces are treated in the interaction in Fourier space without specifying the interaction
potential. The cooling force is explicitly calculated for a regularized and screened potential
which is both of finite range and less singular than the Coulomb interaction at the origin.
Closed expressions are derived for monochromatic electron beams,
which are folded with the velocity distributions of the electrons and ions. The resulting
cooling force is evaluated for anisotropic Maxwell velocity distributions of the electrons and ions.
\end{abstract}

\maketitle

\section{Introduction}
\label{sec:intr}

In most experiments with particle beams a high phase space density is desired. In electron
cooling of ion beams \cite{bud67} this is achieved by mixing the ion beam with a comoving
electron beam which has a very small longitudinal momentum spread. In the rest frame of the
beams the cooling process may be viewed as the stopping of ions in an electron plasma
\cite{sor83,pot90,mes94,men08}. More recently electron cooling has also been used in traps
for precision experiments like CPT--tests with antihydrogen \cite{amo02,gab02} or planned
QED--tests with highly charged ions in HITRAP \cite{qui01}. In these applications the presence
of strong external magnetic fields constitutes a theoretical challenge \cite{der78}, as its
influence on the cooling which the magnetized electrons exert on the ions (antiprotons) is not
so obvious as earlier models might suggest. In the dielectric theory the drag on the
ion is due to the polarization it creates in its wake. This can be either calculated in linear--response
(LR) \cite{wal02,ner00} or numerically by a particle--in--cell (PIC) simulation of
the underlying nonlinear Vlasov--Poisson equation \cite{wal00,mol04}. While the LR requires
cutoffs to exclude hard collisions of close particles the collectivity of the excitation
can be taken into account in both approaches. In the complementary binary collision (BC)
approximation the drag force is accumulated from the velocity transfers in individual collisions.
This has been calculated by scattering statistical ensembles of magnetized electrons from the
ions in the classical trajectory Monte Carlo method (CTMC) \cite{mol04,zwi00,zwi99,zwi02,zwi06,zwi08},
and by treating the Coulomb interaction as a perturbation to the helical motion of the electrons
\cite{toe02,mol03,ners03,ner03,ner09,ner10,ner07}. The observed cooling force $\mathbf{F}(\mathbf{v}_{i})$
on an individual ion is obtained by integrating with respect to
the impact parameter and the electrons velocity distribution. The ion velocity $\mathbf{v}_{i}$
is measured with respect to the center of that distribution. As in electron cooler the electrons
are accelerated from the cathode, their velocity distribution is flattened longitudinally, but
the spread does not vanish. And since the cooling force on slow ions and therefore the cooling process
depends critically on the details of the velocity distribution, a treatment employing a
realistic velocity distribution is desirable.

The purpose of this paper is the application of a second-order perturbative BC model for calculating
the magnetized cooling force on a uniformly moving individual heavy ion as well as on a heavy ion beam.
In previous approaches~\cite{toe02,mol03} three regimes are identified, depending on the relative size
of the cyclotron radius, the distance of the closest approach, and the pitch of the helix. The present
paper is based on our earlier studies in Refs.~\cite{ners03,ner03,ner09,ner10,ner07} where the second-order
energy transfers for individual collisions of electron--ion \cite{ners03,ner03,ner09,ner07}, of any two
identical particles, like e.g.~electron--electron \cite{ner09} and finally of two gyrating arbitrary
charged particles \cite{ner10} have been calculated with the help of an improved BC treatment. This treatment
is -- e.g.~unlike Refs.~\cite{toe02,mol03} -- valid for any strength of the magnetic field. In Sec.~\ref{sec:s2}
we introduce a perturbative binary collision formulation in terms of the binary force acting between an
ion and a magnetized electron, and derive general expressions for the second-order (with respect to the
interaction potential) cooling forces. In contrast to the previous investigations in
Refs.~\cite{ners03,ner03,ner09,ner10,ner07} we here consider the (macroscopic) cooling forces which are
obtained by integrating the binary force of an individual electron--ion interaction with respect to the
impact parameter and the velocity distribution function of electrons. That is, the cooling force for
monoenergetic electrons is folded with an anisotropic velocity distribution which is typical for electron
cooling of ion beams in storage rings, where the velocity spread is much smaller longitudinal than transverse
to the guiding magnetic field. The resulting expressions involve all cyclotron harmonics of the electrons'
helical motion, and are valid for any interaction potential and any strength of the magnetic field and
anisotropy of the velocity distribution of the electron beam. In Sec.~\ref{sec:s2-3} we present explicit
analytic expressions of this second-order cooling force for the specific case of a regularized and screened
interaction potential \cite{kel63,deu77} which is both of finite range and less singular than the Coulomb
interaction at the origin and which includes as limiting cases the Debye (i.e., screened) and the Coulomb
potentials. For comparison of our expressions with previous approaches we consider in Sec.~\ref{sec:s3}
the corresponding asymptotic expressions for large and small ion velocities and strong and vanishing magnetic
fields. The analytical expressions presented in Sec.~\ref{sec:s2-3} are evaluated numerically in Sec.~\ref{sec:s4}
using parameters of the ESR storage ring at GSI~\cite{win96,wink96,win97}. In particular, we compare our
approach with the CTMC simulations and the empirical formula of Parkhomchuk~\cite{par00,park00}. In Sec.~\ref{sec:s5}
we calculate the magnetized cooling force averaged with respect to the ion beam velocity distribution function.
As in Sec.~\ref{sec:s2-3} a similar anisotropic distribution is used for averaging with respect to the ion
velocity distribution. Furthermore, in Sec.~\ref{sec:s5-0} for the resulting cooling force the asymptotic
expressions for large and small ion velocities and strong and vanishing magnetic fields are given. In Sec.~\ref{sec:s5-2}
we compare our approach with the experimental data of the ESR storage ring \cite{win96,wink96,win97}. The
results are summarized and discussed in Sec.~\ref{sec:sum}. In Appendix~\ref{sec:app1} we compare
our asymptotic expressions for the cooling force with those obtained in Ref.~\cite{par85} and demonstrate
that the deviations between both treatments are related to the divergent nature of the bare Coulomb interaction
employed in Ref.~\cite{par85}. The regularization parameter and the screening length involved in the interaction
potential are specified and discussed in Appendices~\ref{sec:app2} and \ref{sec:app3}.

\section{Theoretical model}
\label{sec:s2}

\subsection{Binary collision (BC) formulation}
\label{sec:s2-1}

We consider two point charges with masses $m$, $M$ and charges $-e$, $Ze$, respectively, moving
in a homogeneous magnetic field $\mathbf{B}=B\mathbf{b}$. We assume that the particles interact
with the potential $-Ze\!\!\!/^{2}U(\mathbf{r})$ with $e\!\!\!/^{2}=e^{2}/4\pi \varepsilon _{0}$,
where $\varepsilon _{0}$ is the permittivity of the vacuum and $\mathbf{r}=\mathbf{r}_{1}-\mathbf{r}_{2}$
is the relative coordinate of the colliding particles. For two isolated charged particles this
interaction is given by the Coulomb potential, i.e.~$U_{\mathrm{C}}(\mathbf{r})=1/r$.
In plasma applications $U_{\mathrm{C}}$ is modified by many-body effects and the related screening and turns into an effective interaction. In general, this effective interaction, which is related to the wake field induced by a moving ion, is non-spherically symmetric and depends also on the ion velocity. For any BC treatment, however, this
complicated ion-plasma interaction must be approximated by an effective two particle interaction $U(\mathbf{r})$. This effective interaction $U$ may be modeled by a spherically symmetric
Debye-like screened interaction $U_{\mathrm{D}}(\mathbf{r})=e^{-r/\lambda }/r$ with a screening length $\lambda $, given e.g.~by the Debye screening length $\lambda _{\mathrm{D}}$, see, for example \cite{akh75}, in case of low ion velocities and an
effective velocity dependent screening length $\lambda(v_i)$ for larger ion velocities $v_i$, see \cite{zwi09,zwi02a,zwi99a}. Further details on the choice of the effective interaction $U(\mathbf{r})$ are given in Appendix~\ref{sec:app2}.
To cure problems related to the Coulomb singularity in a classical picture and preventing particles (for $Z>0$)
from falling into the center of these potentials, the screened interaction $U_{\mathrm{D}}$ is replaced
with an effective interaction $U_{\mathrm{R}}$ which is regularized at the origin, taking for example
$U_{\mathrm{R}}(\mathbf{r})=(1-e^{-r/\lambdabar })e^{-r/\lambda }/r$ \cite{kel63,deu77}. Here the use
of this regularized interaction essentially represents an alternative implementation of the standard
(lower) cutoff procedure needed to handle the hard collisions in a classical perturbative approach.
Hence we consider $\lambdabar $ as a given constant or as a function of the classical collision
diameter (see Appendix~\ref{sec:app2}).

In the presence of an external magnetic field, the Lagrangian and the corresponding equations of
particles motion cannot, in general, be separated into parts describing the relative motion and the
motion of the center of mass (cm) \cite{ner07}. However, in the case of heavy ions, i.e. $M\gg m$,
the equations of motion can be simplified by treating the cm velocity $\mathbf{v}_{\rm cm}$ as a
constant and equal to the ion velocity $\mathbf{v}_{i}$, i.e. $\mathbf{v}_{\rm cm}= \mathbf{v}_{i}
=\mathrm{const}$. Then the equation of relative motion turns into
\begin{equation}
{\dot{\mathbf{v}}}(t)+\omega _{c}\left[ \mathbf{v}(t)\times \mathbf{b}\right]
=-\omega _{c}\left[ \mathbf{v}_{i}\times \mathbf{b}\right] -\frac{%
Ze\!\!\!/^{2}}{m}\mathbf{f}\left( \mathbf{r}(t)\right) ,
\label{eq:a1}
\end{equation}%
where $\mathbf{v}(t)={\dot{\mathbf{r}}}(t)=\mathbf{v}_{e}(t)-\mathbf{v}_{i}$ is the relative
electron--ion velocity, $-Ze\!\!\!/^{2}\mathbf{f}(\mathbf{r}(t))$ ($\mathbf{f}=-\partial U/\partial \mathbf{r}$)
is the force exerted by the ion on the electron, $\omega _{c}=eB/m$ is the electron cyclotron frequency.

It is now useful to introduce the velocity correction through relations $\delta \mathbf{v}(t)=%
\mathbf{v}_{e}(t)-\mathbf{v}_{e0}(t)=\mathbf{v}(t)-\mathbf{v}_{0}(t)$, where $\mathbf{v}_{e0}(t)$ and
$\mathbf{v}_{0}(t)$ are the unperturbed electron and relative velocities, respectively, with
$\mathbf{v}_{0}(t)={\dot{\mathbf{r}}}_{0}(t)=\mathbf{v}_{e0}(t)-\mathbf{v}_{i}$,
\begin{eqnarray}
&&\mathbf{r}_{0}(t) =\mathbf{R}_{0}+\mathbf{v}_{r}t+a\left[
\mathbf{u}\sin \left( \omega _{c}t\right) -\left[ \mathbf{b}\times \mathbf{u}%
\right] \cos \left( \omega _{c}t\right) \right] , \label{eq:a2}  \\
&&\delta {\dot{\mathbf{v}}}(t)+\omega _{c}\left[ \delta \mathbf{v}(t)\times
\mathbf{b}\right] =-\frac{Ze\!\!\!/^{2}}{m}\mathbf{f}\left[ \mathbf{r}(t)\right] \label{eq:a3}
\end{eqnarray}%
and $\delta \mathbf{v}(t)\to 0$ at $t\to -\infty $. In Eq.~\eqref{eq:a2} $\mathbf{u}=(\cos \varphi ,\sin
\varphi )$ is the unit vector perpendicular to the magnetic field, the angle $\varphi $ is the initial
phase of the electron's helical motion, $v_{e\parallel }$ and $v_{e\bot }$ (with $v_{e\bot }\geqslant 0$)
are the unperturbed components of the electron velocity parallel and perpendicular to $\mathbf{b}$, respectively,
$\mathbf{v}_{r}=v_{e\parallel }\mathbf{b}-\mathbf{v}_{i}$ is the relative velocity of the
guiding center of the electrons, and $a=v_{e\bot }/\omega _{c}$ is the cyclotron radius.
In Eq.~\eqref{eq:a2}, the variables $\mathbf{u}$ and $\mathbf{R}_{0}$ are independent and are defined by the
initial conditions. In Eq.~\eqref{eq:a3} $\mathbf{r}(t)=\mathbf{r}_{e}(t)-\mathbf{v}_{i}t$ is the ion--electron
relative coordinate. We also introduce the variable $\mathbf{s}=\mathbf{R}_{0\bot }=\mathbf{R}%
_{0}-\mathbf{n}_{r}(\mathbf{n}_{r}\cdot \mathbf{R}_{0})$ which is the component of $\mathbf{R}_{0}$ perpendicular
to the relative velocity vector $\mathbf{v}_{r}$ with $\mathbf{n}_{r}=\mathbf{v}_{r}/v_{r}$. From
Eq.~\eqref{eq:a2} we can see that $\mathbf{s}$ is the distance of closest approach between the ion and the
guiding center of the electron's helical motion.

We seek an approximate solution of Eq.~\eqref{eq:a3} in which the interaction force between the ion and
electrons is considered as a perturbation. Thus we are looking for a solution of Eq.~\eqref{eq:a3} for the
variables $\mathbf{r}$ and $\mathbf{v}$ in a perturbative manner $\mathbf{r}=\mathbf{r}_{0}+\mathbf{r}_{1}+...$,
$\mathbf{v}=\mathbf{v}_{0}+\mathbf{v}_{1}+...$, where $\mathbf{r}_{0}(t),\mathbf{v}_{0}(t)$ are the
unperturbed ion--electron relative coordinate and velocity, respectively, $\mathbf{r}_{n}(t),\mathbf{v}_{n}(t)$
($n=1,2,...$) are the $n$th order perturbations of $\mathbf{r}(t)$ and $\mathbf{v}(t)$, which
are proportional to $Z^{n}$.

The parameter of smallness which justifies such kind of expansion can be read off from a dimensionless form of
the equation of motion Eq.~\eqref{eq:a3} by scaling lengths in units of the screening length $\lambda $,
velocities in units of the initial relative velocity $v_{0}$ and time in units of $\lambda/v_{0}$. In terms
of the scaled quantities, $\widetilde{\mathbf{r}}$, $\delta \widetilde{\mathbf{v}}$, $\widetilde{\mathbf{f}}=
\lambda^{2}\mathbf{f}$, and $\widetilde{\omega}_{c}=\omega_{c}\lambda /v_{0}$, Eq.~\eqref{eq:a3} turns into
\begin{equation}
{\delta \dot{\widetilde{{\mathbf{v}}}}}(t) +
\widetilde{\omega} _{c}\left[\delta \widetilde{\mathbf{v}}(t)\times \mathbf{b}\right]
= - \frac{Ze\!\!\!/^{2}}{m v_{0}^2\lambda } \widetilde{\mathbf{f}}[\widetilde{\mathbf{r}}(t)] .
\label{eq:a3scaled}
\end{equation}%
A perturbative treatment is essentially applicable in cases where  ${|Z|e\!\!\!/^{2}}/{m v_{0}^2\lambda } < 1$, that is,
when the (initial) kinetic energy of relative motion $m v_{0}^{2}/2$ is large compared to the characteristic potential energy
${|Z|e\!\!\!/^{2}}/\lambda$ in a screened Coulomb potential. Or expressed in velocities,
the initial relative velocity $v_{0}$ must exceed the characteristic velocity
$v_{d} =({|Z|e\!\!\!/^{2}}/{m \lambda})^{1/2}$, that is, $v_{d}$ here
demarcates the perturbative from the non-perturbative regime. If this condition is met not only for a single
ion-electron collision but in the average over the electron distribution, e.g.~by replacing $v_{0}$ with the averaged initial ion-electron relative velocity $\langle v_{0} \rangle$, i.e.
\begin{equation}
\langle v_{0} \rangle \gtrsim  v_{d} = \left(\frac{|Z|e\!\!\!/^{2}}{m \lambda}\right)^{1/2}\, ,
\label{eq:deflinregime}
\end{equation}
we are in a regime of weak ion-target, or here, weak ion-electron
coupling, which allows the use of perturbative treatments (besides BC also e.g.~linear-response (LR)).
For nonmagnetized electrons this is discussed in much detail in Refs.~\cite{zwi02a,zwi09}. Even though
the particle trajectories are much more intricate in the presence of an external magnetic field, the given
definitions and demarcations of coupling regimes are basically the same for magnetized electrons. That is, the applicability of a perturbative treatment is essentially related to the charge state $Z$ of the ion and the typical range $\lambda$ of the effective interaction, but not directly on the strength $B$ of the magnetic field. The latter may affect the critical velocity $v_{d}$ only implicitly via a possible change of the effective screening length $\lambda$ with $B$.

The equation for the first--order velocity correction is obtained from Eq.~\eqref{eq:a3} replacing on
the right--hand side the exact relative coordinate $\mathbf{r}(t)$ by $\mathbf{r}_{0}(t)$ with the solutions
$\mathbf{v}_{1}(t)={\dot{\mathbf{r}}}_{1}(t)$ and
\begin{equation}
\mathbf{r}_{1}(t)=\frac{Ze\!\!\!/^{2}}{m}\left\{
-\mathbf{b}\mathcal{Q}_{\parallel }(t)+\Rea \left[ \mathbf{b}\left(
\mathbf{b}\cdot \boldsymbol{\mathcal{Q}}_{\bot }(t)\right) -\boldsymbol{\mathcal{Q}}_{\bot
}(t)+i\left[ \mathbf{b}\times \boldsymbol{\mathcal{Q}}_{\bot }(t)\right] \right] \right\} .
\label{eq:a4}
\end{equation}%
Here we have introduced the following abbreviations
\begin{equation}
\mathcal{Q}_{\parallel }(t)=\int_{-\infty }^{t}\mathbf{b}\cdot
\mathbf{f}\left(\mathbf{r}_{0}(\tau )\right) \left( t-\tau \right) d\tau ,   \quad
\boldsymbol{\mathcal{Q}}_{\bot }(t)=\frac{1}{i\omega _{c}}\int_{-\infty
}^{t}\mathbf{f}\left( \mathbf{r}_{0}(\tau )\right) [e^{i\omega
_{c}\left( t-\tau \right) }-1]d\tau
\label{eq:a5}
\end{equation}%
and have assumed that all corrections vanish at $t\to -\infty $. As will be shown below, Eqs.~\eqref{eq:a2}
and \eqref{eq:a4} completely determine the second-order cooling force on the ion.

\subsection{Second-order cooling forces}
\label{sec:s2-2}

We now consider the interaction process of an individual ion with a
homogeneous electron beam described by a velocity distribution function $f(\mathbf{v}_{e})$
and a density $n_{e}$. We assume that the ion experiences independent
binary collisions (BCs) with the electrons. The total cooling force acting on the ion is then
obtained by multiplying the binary force $Ze\!\!\!/^{2}\mathbf{f}(\mathbf{r}(t))$
by the element of the electron relative flux $n_{e}v_{r}d^{2}\mathbf{s}dt$
(where $\mathbf{s}$ is the impact parameter introduced above which is
perpendicular to the relative velocity $\mathbf{v}_{r}$) and
integrating with respect to time and folding with velocity distribution
of the electrons. The result reads
\begin{equation}
\mathbf{F}\left( \mathbf{v}_{i}\right) =Ze\!\!\!/^{2}n_{e}\int d\mathbf{v}%
_{e}f\left( \mathbf{v}_{e}\right) v_{r}\int
d^{2}\mathbf{s}\int_{-\infty }^{\infty }\mathbf{f}\left(
\mathbf{r}\left( t\right) \right) dt
\label{eq:a6}
\end{equation}%
and is an exact relation for uncorrelated BCs of the ion with electrons. We evaluate this expression
within a systematic perturbative treatment. First, we introduce the two--particle interaction potential
$U(\mathbf{r})$ and the binary force $\mathbf{f}(\mathbf{r})$ is written using Fourier transformation
in space. Furthermore, the factor $e^{i\mathbf{k}\cdot \mathbf{r}(t)}$ in the Fourier transformed binary
force is expanded in a perturbative manner as $e^{i\mathbf{k}\cdot \mathbf{r}(t)}\simeq e^{i\mathbf{k}\cdot%
\mathbf{r}_{0}(t)}[1+i(\mathbf{k}\cdot \mathbf{r}_{1}(t))]$, where $\mathbf{r}_{0}(t)$ and $\mathbf{r}_{1}(t)$
are the unperturbed and the first-order corrected relative coordinates, Eqs.~\eqref{eq:a2} and \eqref{eq:a4},
respectively. Thus the binary force within second-order perturbative treatment turns into
\begin{equation}
\mathbf{f}\left(\mathbf{r}\left( t\right) \right) = -i
\int d\mathbf{k}U\left( \mathbf{k} \right) \mathbf{k}%
\ e^{i\mathbf{k}\cdot \mathbf{r}\left( t\right) } \simeq -i
\int d\mathbf{k}U\left( \mathbf{k} \right) \mathbf{k}%
\left[ 1 + i\left(\mathbf{k}\cdot \mathbf{r}_{1}(t)\right)\right]
e^{i\mathbf{k}\cdot \mathbf{r}_{0}\left( t\right) } .
\label{eq:xxx}
\end{equation}%
The first and the second terms in the last part of Eq.~\eqref{eq:xxx} correspond to the first- ($\mathbf{f}_{1}$)
and the second-order ($\mathbf{f}_{2}$) binary forces, respectively. We consider only the second-order binary
force $\mathbf{f}_{2}$ and the corresponding force $\mathbf{F}_{2}$ with respect to the binary interaction since
the averaged first-order force $\mathbf{F}_{1}$ (related to $\mathbf{f}_{1}$) vanishes due to symmetry reasons
\cite{ner03,ners03,ner07,ner09,ner10}. Within the second-order perturbative treatment the cooling force can be
represented as:
\begin{equation}
\mathbf{F}_{2} =Ze\!\!\!/^{2}n_{e}\int d\mathbf{v}_{e}f\left( \mathbf{v}%
_{e}\right) v_{r}\int d^{2}\mathbf{s}\int d\mathbf{k}U\left( \mathbf{k}%
\right) \mathbf{k}\int_{-\infty }^{\infty }\left( \mathbf{k}\cdot \mathbf{r}%
_{1}(t)\right) e^{i\mathbf{k}\cdot \mathbf{r}_{0}\left( t\right) }dt .
\label{eq:a10}
\end{equation}%
From Eq.~\eqref{eq:a10} it is seen that the second-order cooling force, $\mathbf{F}_{2}$, is proportional to $Z^{2}$.

Substituting Eqs.~\eqref{eq:a4} and \eqref{eq:a5} into Eq.~\eqref{eq:a10} and writing the binary force in
expression~\eqref{eq:a5} in terms of Fourier transformed potential results in
\begin{eqnarray}
&&\mathbf{F}_{2} =\frac{iZ^{2}e\!\!\!/^{4}n_{e}}{m}\int d\mathbf{v}%
_{e}f\left( \mathbf{v}_{e}\right) v_{r}\int d^{2}\mathbf{s}\int d\mathbf{k}d%
\mathbf{k}^{\prime }U\left( \mathbf{k}\right) U\left(\mathbf{k}^{\prime
}\right) \mathbf{k}\int_{-\infty }^{\infty }e^{i\mathbf{k}\cdot \mathbf{r}%
_{0}\left( t\right) }dt\int_{-\infty }^{t}e^{i\mathbf{k}^{\prime }\cdot
\mathbf{r}_{0}\left( \tau \right) }d\tau  \label{eq:a11}  \\
&&\times \left\{ g_{0} \left( t-\tau \right) +\frac{g_{1}}{\omega _{c}}
\sin \left( \omega _{c}(t-\tau )\right)
-\frac{g_{2}}{\omega _{c}} \left[ 1-\cos \left(
\omega _{c}(t-\tau )\right) \right] \right\} ,  \nonumber
\end{eqnarray}%
where $g_{0}=(\mathbf{k}\cdot \mathbf{b}) (\mathbf{k} ^{\prime }\cdot \mathbf{b})$, $g_{1}=(\mathbf{k}\cdot \mathbf{k}%
^{\prime }) -(\mathbf{k}\cdot \mathbf{b}) (\mathbf{k} ^{\prime }\cdot \mathbf{b})$, $g_{2}=(\mathbf{k}\cdot%
[\mathbf{k}^{\prime }\times \mathbf{b}])$. The time--integral in Eq.~\eqref{eq:a11} can be performed using the Fourier
series expansion of the exponential function $e^{iz\sin (\omega t)}=\sum_{n=-\infty}^{\infty} J_{n}(z) e^{in\omega t}$,
where $J_{n}$ are the Bessel functions of the $n$th order (see, e.g., Ref.~\cite{gra80}). This yields
\begin{eqnarray}
&&\mathbf{F}_{2} =\frac{2\pi iZ^{2}e\!\!\!/^{4}n_{e}}{m}\int d\mathbf{v}%
_{e}f\left( \mathbf{v}_{e}\right) v_{r}\int d^{2}\mathbf{s}\int d\mathbf{k}d%
\mathbf{k}^{\prime }U\left( \mathbf{k}\right) U\left(
\mathbf{k}^{\prime }\right) \mathbf{k}e^{i\left(
\mathbf{k}+\mathbf{k}^{\prime }\right) \cdot \mathbf{R}_{0}}  \nonumber \\
&&\times \sum_{n,m=-\infty }^{\infty }e^{i\left( n+m\right)
\varphi }e^{-in\theta -im\theta ^{\prime }}J_{n}\left( k_{\bot}a\right) J_{m}\left(
k_{\bot }^{\prime }a\right) \delta \left( \zeta _{n}(\mathbf{k})+\zeta _{m}(%
\mathbf{k}^{\prime })\right)  \label{eq:a13} \\
&&\times \left\{ -\frac{g_{0}}{\left[ \zeta _{m}(\mathbf{k}^{\prime })-i0%
\right] ^{2}}+\frac{g_{1}}{2\omega _{c}}\left[ \frac{1}{\zeta _{m+1}(\mathbf{%
k}^{\prime })-i0}-\frac{1}{\zeta _{m-1}(\mathbf{k}^{\prime})-i0}\right] \right.  \nonumber \\
&&\left. +\frac{ig_{2}}{2\omega _{c}}\left[ \frac{2}{\zeta _{m}(\mathbf{k}%
^{\prime })-i0}-\frac{1}{\zeta _{m+1}(\mathbf{k}^{\prime })-i0}-\frac{1}{%
\zeta _{m-1}(\mathbf{k}^{\prime })-i0}\right] \right\} .  \nonumber
\end{eqnarray}%
Here $\tan \theta =k_{y}/k_{x}$, $k_{\parallel }=(\mathbf{k}\cdot \mathbf{b})$ and $k_{\bot }$ are the components of
$\mathbf{k}$ parallel and transverse to $\mathbf{b}$, respectively, $\zeta _{n}(\mathbf{k})%
=n\omega _{c}+\mathbf{k}\cdot \mathbf{v}_{r}$, and $\varphi $ is the initial phase of
the electron as defined in the previous Section. Note that expression \eqref{eq:a13} involves
all cyclotron harmonics.

Next, we integrate with respect to the initial phase $\varphi $ and impact parameter $\mathbf{s}$.
For that purpose we recall that the volume element $d\mathbf{v}_{e}$ can be represented in cylindrical
coordinates as $d\mathbf{v}_{e}=dv_{e\parallel }v_{e\bot }dv_{e\bot }d\varphi $, where $v_{e\parallel }$
and $v_{e\bot }$ are the electron velocity components parallel and transverse to $\mathbf{b}$, respectively.
The $\mathbf{s}$--integration is enabled by using the relation $e^{i\mathbf{k}\cdot \mathbf{R}_{0}}=e^{i\kappa
_{\parallel }R_{0\parallel }}e^{i\boldsymbol{\kappa }_{\bot }\cdot \mathbf{s}}$, where $\kappa _{\parallel }=
(\mathbf{k}\cdot \mathbf{n}_{r})$, $\boldsymbol{\kappa }_{\bot }=\mathbf{k}-\mathbf{n}_{r}(\mathbf{k}\cdot
\mathbf{n}_{r})$, i.e. the component of $\mathbf{k}$ parallel and transverse to $\mathbf{n}_{r}$.
Performing now the $\varphi $ and $\mathbf{s}$--integrations results in
\begin{eqnarray}
&&\mathbf{F}_{2} =-\frac{\left( 2\pi \right) ^{5}Z^{2}e\!\!\!/^{4}n_{e}}{%
2m}\int_{-\infty }^{\infty }dv_{e\parallel
}\int_{0}^{\infty }f\left( v_{e\parallel },v_{e\bot }\right) v_{e\bot }dv_{e\bot }\int d%
\mathbf{k}\left\vert U\left( \mathbf{k}\right) \right\vert ^{2}\mathbf{k} \label{eq:a15} \\
&&\times \sum_{n=-\infty }^{\infty }J_{n}^{2}\left( k_{\bot }a\right)
\left\{ k_{\parallel }^{2}\delta ^{\prime }\left( \zeta _{n}(\mathbf{k}%
)\right) +\frac{k_{\bot }^{2}}{2\omega _{c}}\left[ \delta \left(\zeta
_{n+1}(\mathbf{k})\right) -\delta \left( \zeta _{n-1}(\mathbf{k})\right) %
\right] \right\} ,  \nonumber
\end{eqnarray}%
where the prime indicates the derivative with respect to the argument. For deriving Eq.~\eqref{eq:a15} we
assumed an axially symmetric velocity distribution $f(\mathbf{v}_{e})=f(v_{e\parallel },v_{e\bot })$ and
used $\delta (\kappa _{\parallel })\delta (\boldsymbol{\kappa }_{\bot}) =\delta (\mathbf{k})$.

The $n$--summation in Eq.~\eqref{eq:a15} can be done using the summation formula for the Bessel functions
\cite{gra80}. We then obtain
\begin{eqnarray}
&&\mathbf{F}_{2} =\frac{\left( 2\pi \right) ^{4}Z^{2} e\!\!\!/^{4} n_{e}}{%
m}\int_{-\infty }^{\infty }dv_{e\parallel }\int_{0}^{\infty
}f\left( v_{e\parallel },v_{e\bot }\right) v_{e\bot }dv_{e\bot }\int d%
\mathbf{k}\left\vert U\left( \mathbf{k}\right) \right\vert ^{2}\mathbf{k} \label{eq:a16} \\
&&\times \int_{0}^{\infty }\left[ k_{\parallel }^{2}+k_{\bot
}^{2}\frac{\sin \left( \omega _{c}t\right) }{\omega _{c}t}\right] J_{0}\left( 2k_{\bot
}a\sin \frac{\omega _{c}t}{2}\right) \sin \left( \mathbf{k}\cdot \mathbf{v}_{r}t\right) tdt .  \nonumber
\end{eqnarray}%
This is a general expression for the magnetized cooling force acting on an individual ion. It has been derived within
second-order perturbation theory but without any restriction on the strength of the magnetic field $B$. The limiting
cases of Eq.~\eqref{eq:a16} at vanishing $B$ and in the presence of an infinitely strong magnetic field are briefly
studied in Sec.~\ref{sec:s3-1} (see also Appendix~\ref{sec:app1}).

\subsection{Cooling force for a regularized and screened Coulomb potential}
\label{sec:s2-3}

In electron cooling of ion beams the velocity distribution of the electrons is anisotropic which is a typical
situation for electron coolers. It is usually modeled by a two--temperature anisotropic Maxwell distribution
with different temperatures for the longitudinal and transverse degrees of freedom. The velocity distribution
relevant for the averaging in Eq.~\eqref{eq:a16} is thus given by
\begin{equation}
f\left( v_{e\parallel },v_{e\bot }\right) =\frac{1}{\left( 2\pi
\right) ^{3/2}v_{\mathrm{th}\bot }^{2}v_{\mathrm{th}\parallel }}e^{-v_{e\bot
}^{2}/2v_{\mathrm{th}\bot }^{2}}e^{-v^{2}_{e\parallel }/2v_{\mathrm{th}\parallel }^{2}} ,
\label{eq:a17}
\end{equation}%
where the thermal velocities are related to electron temperatures by $v_{\mathrm{th}\bot }^{2}=T_{\bot }/m$, $v_{\mathrm{th}%
\parallel}^{2}=T_{\parallel }/m$ (here the temperatures are measured in energy units).
In this case the transverse ($\mathbf{F}_{\bot }=\mathbf{F}-\mathbf{b}F_{\parallel }$) and longitudinal
($F_{\parallel }=\mathbf{b}\cdot \mathbf{F}$) components of the cooling force \eqref{eq:a16} with
Eq.~\eqref{eq:a17} (we dropped the index 2 in $\mathbf{F}_{2}$ for simplicity), after velocity
integrations (see Ref.~\cite{gra80}) can be represented in the forms
\begin{eqnarray}
&&\left\{
\begin{array}{c}
F_{\bot }( \mathbf{v}_{i}) \\
F_{\parallel }(\mathbf{v}_{i})%
\end{array}%
\right\} =-\frac{8Z^{2}e\!\!\!/^{4}n_{e}}{m\omega
_{c}^{2}}\frac{\left( 2\pi \right) ^{4}}{4}\int_{0}^{\infty
}dk_{\parallel }\int_{0}^{\infty
}U^{2}(k) k_{\bot }dk_{\bot }  \label{eq:a18} \\
&&\times \int_{0}^{\infty }e^{-\frac{t^{2}}{2}k_{\parallel
}^{2}a_{\parallel }^{2}}e^{-k_{\bot }^{2}a_{\bot }^{2}\left(
1-\cos t\right) }\left( k_{\parallel }^{2}+k_{\bot }^{2}\frac{\sin
t}{t}\right) \left\{
\begin{array}{c}
k_{\bot }\cos \left( k_{\parallel
}a_{i\parallel }t\right) J_{1}\left( k_{\bot }a_{i\bot }t\right) \\
k_{\parallel }\sin \left( k_{\parallel }a_{i\parallel }t\right)
J_{0}\left(
k_{\bot }a_{i\bot }t\right)%
\end{array}%
\right\} tdt   \nonumber
\end{eqnarray}%
with $\mathbf{F}_{\bot }( \mathbf{v}_{i})=\frac{\mathbf{v}_{i\bot}}{v_{i\bot }} F_{\bot }( \mathbf{v}_{i})$.
Here we have assumed a spherically symmetric potential $U(\mathbf{k})=U(k)$ and have introduced the thermal
cyclotron radii of the electrons $a_{\bot} =v_{\mathrm{th}\bot }/\omega _{c}$, $a_{\parallel} =v_{\mathrm{th}\parallel}
/\omega _{c}$, and $a_{i\bot }=v_{i\bot}/\omega _{c}$, $a_{i\parallel }=v_{i\parallel }/\omega _{c}$. In general
the cooling force is thus anisotropic with respect to the ion velocity $\mathbf{v}_{i}$.

For the Coulomb interaction $U(k)=U_{\mathrm{C}}(k)$, the full two--dimensional integration over the $\mathbf{s}$--space
results in a logarithmic divergence of the $\mathbf{k}$-integration in Eqs.~\eqref{eq:a15} and \eqref{eq:a16}. To cure this,
cutoff parameters $k_{\min }$ and $k_{\max }$ must be introduced, see, e.g., Refs.~\cite{ner03,ners03,ner07} for details.
Instead of doing so, we here employ the regularized screened potential $U(\mathbf{r})=U_{\mathrm{R}}(r)$ introduced in
Sec.~\ref{sec:s2-1} with the Fourier transform
\begin{equation}
U_{\mathrm{R}}(k)=\frac{2}{\left( 2\pi \right) ^{2}}\left( \frac{1}{%
k^{2}+\lambda ^{-2}}-\frac{1}{k^{2}+d^{-2}}\right) ,
\label{eq:a19}
\end{equation}%
where $d^{-1}=\lambda ^{-1}+\lambdabar ^{-1}$.

Substituting the interaction potential~\eqref{eq:a19} into Eq.~\eqref{eq:a18} and performing the
$k_{\parallel}$--integration we arrive, after lengthly but straightforward calculations, at
\begin{eqnarray}
&&F_{\parallel }(\mathbf{v}_{i}) =-\frac{4\sqrt{\pi }Z^{2}e\!\!\!/^{4}n_{e}}{%
mv_{\mathrm{th}\parallel }^{2}}\upsilon _{\parallel }\int_{0}^{\infty }\frac{%
dt}{t}\int_{0}^{1}d\zeta \Phi \left( \psi (t,\zeta )\right) \exp \left[
-\upsilon _{\parallel }^{2}\zeta ^{2}-\frac{\upsilon _{\perp }^{2}\zeta ^{2}%
}{G(t,\zeta )}\right]  \label{eq:y1}  \\
&&\times \frac{\zeta ^{2}\left( 1-\zeta ^{2}\right) }{G(t,\zeta)} \left\{ 3-2\upsilon _{\parallel }^{2}\zeta ^{2}+\frac{2}{G(t,\zeta )%
}\left[ 1-\frac{\upsilon _{\perp }^{2}\zeta ^{2}}{G(t,\zeta )}\right] \frac{%
\sin (\alpha t)}{\alpha t}\right\} ,  \nonumber \\
&&F_{\bot }(\mathbf{v}_{i}) =-\frac{4\sqrt{\pi }Z^{2}e\!\!\!/^{4}n_{e}}{mv_{%
\mathrm{th}\parallel }^{2}}\upsilon _{\perp }\int_{0}^{\infty }\frac{dt}{t}%
\int_{0}^{1}d\zeta \Phi \left( \psi (t,\zeta )\right) \exp \left[ -\upsilon
_{\parallel }^{2}\zeta ^{2}-\frac{\upsilon _{\perp }^{2}\zeta ^{2}}{G(t,\zeta )}\right]   \label{eq:y2}  \\
&&\times \frac{\zeta ^{2}\left( 1-\zeta ^{2}\right) }{G^{2}(t,\zeta )} \left\{ 1-2\upsilon _{\parallel }^{2}\zeta ^{2}+\frac{2}{G(t,\zeta )%
}\left[ 2-\frac{\upsilon _{\perp }^{2}\zeta ^{2}}{G(t,\zeta )}\right] \frac{%
\sin (\alpha t)}{\alpha t}\right\} ,  \nonumber
\end{eqnarray}%
where we have introduced the dimensionless quantities $\upsilon _{\parallel }=v_{i\parallel }/\sqrt{2}
v_{\mathrm{th}\parallel }$, $\upsilon _{\perp }=v_{i\bot }/\sqrt{2}v_{\mathrm{th}\parallel}$, $\alpha =\omega _{c}
\lambda /v_{\mathrm{th}\parallel }$, and $\tau =T_{\bot }/T_{\parallel }$ is the anisotropy parameter of the electron
beam. Here $\psi (t,\zeta )=(t^{2}/2)(1-\zeta ^{2})/\zeta ^{2}$, $G(t,\zeta )=\tau \Theta (t)\zeta ^{2}+1-\zeta ^{2}$,
$\Theta (t) =\left( \frac{2}{\alpha t}\sin \frac{\alpha t}{2}\right) ^{2}$, and
\begin{equation}
\Phi (z) =e^{-z}+e^{-\varkappa ^{2}z}-\frac{2}{\varkappa ^{2}-1}%
\frac{1}{z}\left( e^{-z}-e^{-\varkappa ^{2}z}\right) ,
\label{eq:axx}
\end{equation}
where $\varkappa =\lambda /d=1+ \lambda /\lambdabar $. Equations~\eqref{eq:y1} and \eqref{eq:y2} for the parallel and
transversal components of the drag force, respectively, are the main results of this paper. In the next section we
compare systematically these expressions as well as general Eq.~\eqref{eq:a16} with previous approaches.

\section{Comparison with previous approaches}
\label{sec:s3}

Previous theoretical expressions for the cooling force which have been extensively discussed by electron cooling
community (see, e.g., Refs.~\cite{der78,men08} for a review) basically concern the two limiting cases
of vanishing and infinitely strong magnetic fields. We therefore consider our previously presented approach in some
detail for these two cases, first for arbitrary interactions $U(\mathbf{k})$ and electron
distributions $f(\mathbf{v}_{e})$ as given by Eq.~\eqref{eq:a16} and later for the specific situation of the regularized
interaction \eqref{eq:a19} and the velocity distribution \eqref{eq:a17} as given by Eqs.~\eqref{eq:y1} and \eqref{eq:y2}.

\subsection{Cooling force Eq.~(\ref{eq:a16}) at vanishing and infinitely strong magnetic fields}
\label{sec:s3-1}

For $B\to 0$, i.e.~at vanishing magnetic field, $\sin (\omega _{c}t)/(\omega _{c}t)\to 1$ and the argument of the Bessel
function in Eq.~\eqref{eq:a16} should be replaced by $k_{\bot }v_{e\bot}t$. Then, denoting the second-order force at
vanishing magnetic field as $\mathbf{F}_{0}$ and using an integral representation of the Bessel function $J_{0}$, one obtains
\begin{eqnarray}
&&\mathbf{F}_{0}(\mathbf{v}_{i})= -\frac{\left( 2\pi \right) ^{3}Z^{2}e\!\!\!/^{4}n_{e}}{m}\int f( \mathbf{v}%
_{e}) d\mathbf{v}_{e}\int d\mathbf{k}\left\vert U\left( \mathbf{k}%
\right) \right\vert ^{2}k^{2}\mathbf{k} \label{eq:x1}  \\
&&\times \frac{\partial }{\partial \omega }%
\int_{0}^{\infty } J_0 \left( \mathbf{k}_{\bot }\cdot \mathbf{v}_{e\bot
}t\right) \cos \left( \omega t\right) dt
=\frac{4\pi Z^{2}e\!\!\!/^{4}n_{e}}{m}\frac{\partial }{\partial \mathbf{v}_{i}}\int
G_{0}(\bar{\mathbf{v}}_{r}) f(\mathbf{v}_{e}) d\mathbf{v}_{e}  \nonumber
\end{eqnarray}
with
\begin{equation}
G_{0}(\bar{\mathbf{v}}_{r}) =\frac{\left( 2\pi \right) ^{3}}{4}\int \left\vert
U\left( \mathbf{k}\right) \right\vert ^{2}\delta \left( \mathbf{k}\cdot
\bar{\mathbf{v}}_{r}\right) k^{2}d\mathbf{k} .
\label{eq:x2}
\end{equation}%
Here $\omega =\mathbf{k}\cdot \mathbf{v}_{r}$, $\bar{\mathbf{v}}_{r}=\mathbf{v}_{r}+\mathbf{v}_{e\perp}=\mathbf{v}
_{e}-\mathbf{v}_{i}$, $\mathbf{v}_{e}$ and $\bar{\mathbf{v}}_{r}$ are the three--dimensional electron and the
ion--electron relative velocities, respectively. The other quantities in Eqs.~\eqref{eq:x1} and \eqref{eq:x2}
have been introduced in Sec.~\ref{sec:s2}. In particular, assuming spherically symmetric potential with $U(\mathbf{k})
=U(k)$, from Eq.~\eqref{eq:x2} it is straightforward to obtain $G_{0}(\bar{\mathbf{v}}_{r}) =G_{0}(\bar{v}_{r}) =
(1/\bar{v}_{r})\mathcal{U}$ and thus
\begin{equation}
\mathbf{F}_{0}(\mathbf{v}_{i})=
\frac{4\pi Z^{2}e\!\!\!/^{4}n_{e}}{m} \ \mathcal{U} \int
\frac{\mathbf{v}_{e}-\mathbf{v}_{i} }{|\mathbf{v}_{e}-\mathbf{v}_{i}|^3}
 f(\mathbf{v}_{e}) d\mathbf{v}_{e} ,
\label{eq:x21}
\end{equation}
where $\mathcal{U}$ is the generalized Coulomb logarithm,
\begin{equation}
\mathcal{U}  =\frac{(2\pi )^{4}}{4}\int_{0}^{\infty }U^{2}(k) k^{3}dk .
\label{eq:x3}
\end{equation}

Employing the regularized and screened potential $U(k)$ given by Eq.~\eqref{eq:a19}, the generalized Coulomb
logarithm is $\mathcal{U}=\mathcal{U}_{\mathrm{R}} =\Lambda (\varkappa)$ (see also Refs.~\cite{ner03,ner09,ner10,ner07}),
where
\begin{equation}
\Lambda (\varkappa )=\frac{\varkappa ^{2}+1}{\varkappa ^{2}-1} \ln \varkappa -1 .
\label{eq:a25}
\end{equation}%
Taking the bare Coulomb interaction with $U(k)=U_{\mathrm{C}}
(k) \sim 1/k^{2}$, Eq.~\eqref{eq:x3} diverges logarithmically at $k\to 0$ and $k\to \infty$ and two cutoffs $k_{\min }
= 1/r_{\max }$ and $k_{\max } =1/r_{\min}$ must be introduced as discussed in Sec.~\ref{sec:s2-3}. In this case the
generalized Coulomb logarithm takes the standard form $\mathcal{U}=\mathcal{U}_{\mathrm{C}} = \ln (k_{\max}/k_{\min})
= \ln (r_{\max}/r_{\min})$.

While the cooling force \eqref{eq:x21} is even at vanishing magnetic field anisotropic due to the anisotropic velocity distribution of the electrons, the asymptotic expression of
\eqref{eq:x21} at high ion velocities is isotropic and
can be easily derived by replacing $\bar{v}_{r} = |\mathbf{v}_{e}-\mathbf{v}_{i}|$ with the ion velocity $\bar{v}_{r}
\simeq v_{i}$ which results in
\begin{equation}
\mathbf{F}_{0}(\mathbf{v}_{i}) \simeq -\frac{4\pi Z^{2}e\!\!\!/^{4}n_{e}}{m v_{i}^{2}} \ \mathcal{U} \
\frac{\mathbf{v}_{i} }{v_{i}} .
\label{eq:x22}
\end{equation}

At an infinitely strong magnetic field $B\to \infty$ the term in Eq.~\eqref{eq:a16} proportional to $k_{\bot }^{2}$
and the argument of the Bessel function vanish since the cyclotron radius $a\to 0$. In this limit, denoting the force
as $\mathbf{F}_{\infty}(\mathbf{v}_{i})$, we arrive at
\begin{equation}
\mathbf{F}_{\infty}(\mathbf{v}_{i}) =\frac{2\pi Z^{2}e\!\!\!/^{4}n_{e}}{m}\frac{\partial }{\partial \mathbf{v}_{i}}%
\int G_{\infty }(\mathbf{v}_{r}) f_{e}(\mathbf{v}_{e}) d\mathbf{v}_{e} ,
\label{eq:x4}
\end{equation}
where
\begin{equation}
G_{\infty }(\mathbf{v}_{r}) =\frac{\left( 2\pi \right) ^{3}}{2}\int \left\vert U\left( \mathbf{k}\right)
\right\vert ^{2}\delta \left( \mathbf{k}\cdot \mathbf{v}_{r}\right) k_{\parallel }^{2}d\mathbf{k} .
\label{eq:x5}
\end{equation}
Again, assuming a spherically symmetric interaction potential from Eq.~\eqref{eq:x5} we obtain $G_{\infty }(\mathbf{v}_{r})
=(v_{i\perp}^{2}/v_{r}^{3})\mathcal{U}$, where $v_{i\perp}$ is the component of the ion velocity perpendicular to the
magnetic field and $\mathcal{U}$ is given by Eq.~\eqref{eq:x3}. Inserting $G_{\infty }(\mathbf{v}_{r})=(v_{i\perp}^{2}/v_{r}^{3})\mathcal{U}$ into Eq.~\eqref{eq:x4} then provides the two components of the cooling force
\begin{eqnarray}
&&F_{\infty \parallel }(\mathbf{v}_{i}) = \frac{6\pi Z^{2}e\!\!\!/^{4}n_{e}}{m}
\ \mathcal{U} \int \frac{v_{i\perp}^{2}v_{r\parallel }}{v_{r}^{5}} f_{e}(\mathbf{v}_{e}) d\mathbf{v}_{e} ,  \label{eq:x41}  \\
&&F_{\infty \bot }(\mathbf{v}_{i}) = \frac{2\pi Z^{2}e\!\!\!/^{4}n_{e}}{m}
\ \mathcal{U} \int \frac{ v_{i\bot } (2 v_{r\parallel }^{2} - v_{i\perp }^{2}) }{v_{r}^{5}} f_{e}(\mathbf{v}_{e}) d\mathbf{v}_{e} ,  \label{eq:x42}
\end{eqnarray}
where $v_{r\parallel }=v_{e\parallel }-v_{i\parallel }$. The corresponding high--velocity asymptotic expressions,
replacing now $\mathbf{v}_{r}$ by $-\mathbf{v}_{i}$, are given by
\begin{eqnarray}
&&F_{\infty \parallel }(\mathbf{v}_{i}) \simeq  - \frac{6\pi Z^{2}e\!\!\!/^{4}n_{e}}{m v_{i}^{2}}
\ \mathcal{U} \ \frac{v_{i\perp}^{2}v_{i\parallel }}{v_{i}^{3}} ,  \label{eq:x43}  \\
&&F_{\infty \bot }(\mathbf{v}_{i}) \simeq \frac{2\pi Z^{2}e\!\!\!/^{4}n_{e}}{m v_{i}^{2}}
\ \mathcal{U} \ \frac{v_{i\bot} (2 v_{i\parallel }^{2} - v_{i\perp }^{2}) }{v_{i}^{3}} .  \label{eq:x44}
\end{eqnarray}
Note that Eqs.~\eqref{eq:x43} and \eqref{eq:x44} can be also obtained from Eqs.~\eqref{eq:x41} and \eqref{eq:x42},
respectively, in the case of a completely flattened distribution function of the electrons in the limit $T_{\parallel} \to 0$ when
the distribution function~\eqref{eq:a17} is given by a delta-function with respect to $v_{e\parallel}$.

Equations~\eqref{eq:x21} and \eqref{eq:x41}, \eqref{eq:x42} and their asymptotic expressions for high--velocities (Eqs.~\eqref{eq:x22}
and \eqref{eq:x43}, \eqref{eq:x44}, respectively) assuming the Coulomb interaction potential with $\mathcal{U}=\mathcal{U}_{\mathrm{C}}$
yield the cooling forces obtained previously in the cases of vanishing and infinitely strong magnetic fields, see e.g.~\cite{der78},
respectively. Equations~\eqref{eq:x1} and \eqref{eq:x4} with a regularized interaction potential thus agree with the similar results
derived by Derbenev and Skrinsky in Ref.~\cite{der78} except for the different Coulomb logarithms $\mathcal{U}$. A more detailed
discussion and comparison of $\mathcal{U}_{\mathrm{R}}=\Lambda (\varkappa )$ given by Eq.~\eqref{eq:a25} and the standard Coulomb logarithm
$\mathcal{U}_{\mathrm{C}}= \ln (r_{\max}/r_{\min})$ can be found in Appendix~\ref{sec:app2}. We like to emphasize here that the Coulomb
logarithm $\mathcal{U}_{\mathrm{R}}$ for the regularized interaction potential has the advantage to allow closed analytic expressions
and converging integrals and avoids any introduction of lower and upper cutoffs 'by hand' in order to restrict the domains of integration.
Moreover, employing the bare Coulomb interaction may, as pointed out by Parkhomchuk \cite{par85}, result in asymptotic expressions which
essentially different from Eqs.~\eqref{eq:x41}--\eqref{eq:x44}. In Appendix~\ref{sec:app1} we show how this is related to the divergent
nature of the bare Coulomb interaction.

\subsection{Some limiting cases of Eqs.~(\ref{eq:y1}) and (\ref{eq:y2})}
\label{sec:s3-2}

More specifically we next discuss some asymptotic regimes of the cooling forces Eqs.~\eqref{eq:y1} and \eqref{eq:y2} when assuming
the regularized interaction \eqref{eq:a19} and the two-temperature velocity distribution \eqref{eq:a17}. In the high--velocity limit
where $v_{i}>(\omega _{c}\lambda ,v_{\mathrm{th}\parallel ;\perp})$ only small $t$ contribute to the cooling forces \eqref{eq:y1}
and \eqref{eq:y2} due to the short time response of the electrons to the moving fast ion. In this limit we have $\sin(\alpha t)/\alpha
t \to 1$ and $G(t,\zeta )\to \tau \zeta^{2}+1-\zeta^{2}$. The remaining $t$--integration can be performed explicitly. This integral
is given by
\begin{eqnarray}
&&\int_{0}^{\infty} \frac{dt}{t}\Phi\left(\psi (t,\zeta )\right) = \lim\limits_{\varepsilon \to \, 0^+} \frac{1}{2}
\int_{\varepsilon}^{\infty} \frac{dz}{z}\Phi\left(z\right) \label{eq:ap1} \\
&&= \lim\limits_{\varepsilon \to \, 0^+} \frac{\varkappa^{2} +1}{2(\varkappa^{2} -1)} [E_{1}(\varepsilon )
-E_{1}(\varkappa^{2} \varepsilon )] -1 \equiv \Lambda (\varkappa) . \nonumber
\end{eqnarray}
Here $z = (t^{2}/2)(1/\zeta^{2}-1)$ was introduced as new variable of integration, the function $\Phi (z)$ is determined by
Eq.~\eqref{eq:axx}, $E_{1}(z) =-\Ei (-z)$ is the exponential integral which behaves at small argument ($z\to 0$) as $E_{1}(z) \simeq
\ln (1/z)-\gamma$ \cite{gra80}, where $\gamma$ is the Euler's constant and $\Lambda(\varkappa)$ is the generalized Coulomb logarithm
Eq.~\eqref{eq:a25}. The remaining expressions do not depend on the magnetic field, i.e.~$\omega _{c}$, as natural consequence of the short
time response of the magnetized electrons. In fact, $\sin(\alpha t)/\alpha t \to 1$ and $G(t,\zeta )\to \tau \zeta^{2}+1-\zeta^{2}$ and
the related $t$--integration \eqref{eq:ap1} are also valid for vanishing magnetic field $\alpha \to 0$. Changing now in the remaining
$\zeta$--integrations the variable $\zeta \to \zeta/(\zeta^{2}+\tau \left( 1-\zeta ^{2}\right))^{1/2}$ turns Eqs.~\eqref{eq:y1} and
\eqref{eq:y2}, after some integration by parts, into
\begin{equation}
F_{\parallel ;\bot }(\mathbf{v}_{i}) =-\frac{8\sqrt{\pi }Z^{2}e\!\!\!/^{4}n_{e}}{mv_{%
\mathrm{th}\parallel ;\perp}^{2}}\Lambda (\varkappa )\upsilon _{\parallel ;\perp }\int_{0}^{1}%
\exp \left[ -\frac{\upsilon _{\parallel }^{2}\zeta ^{2}}{\zeta ^{2}+\tau \left( 1-\zeta ^{2}\right) }-\frac{\upsilon _{\perp
}^{2}\zeta ^{2}}{\tau }\right] \frac{\zeta ^{2}d\zeta }{\left[ \zeta ^{2}+\tau \left( 1-\zeta ^{2}\right) \right] ^{q}} ,
\label{eq:z1}
\end{equation}
where $q =3/2$ and $q =1/2$ for $F_{\parallel} (\mathbf{v}_{i})$ and $F_{\bot} (\mathbf{v}_{i})$, respectively. Here
again the scaled ion velocities $\upsilon _{\parallel }=v_{i\parallel }/\sqrt{2} v_{\mathrm{th}\parallel }$ and
$\upsilon _{\perp }=v_{i\bot }/\sqrt{2}v_{\mathrm{th}\parallel}$ have been used. The cooling forces \eqref{eq:z1}
are anisotropic with respect to the ion velocity $\mathbf{v}_{i}$ due to the anisotropic velocity distribution \eqref{eq:a17}
of the electrons, and they represent the two limiting cases of Eqs.~\eqref{eq:y1} and \eqref{eq:y2}, namely high-velocities
at arbitrary magnetic field and arbitrary velocities at vanishing field. Of course, expression \eqref{eq:z1} can be
also obtained by performing the remaining integration in the nonmagnetized cooling force \eqref{eq:x21} using the
anisotropic velocity distribution \eqref{eq:a17} and $\mathcal{U} = \Lambda (\varkappa )$.

The cooling forces in \eqref{eq:z1} are additionally simplified when the transverse thermal velocity spread
of the electrons $v_{\mathrm{th}\perp}$ is much larger than the longitudinal one $v_{\mathrm{th}\parallel}$ (i.e.~
$T_{\perp }\gg T_{\parallel }$ or $\tau \gg 1$) which is a typical situation for electron coolers. In this case we have
\begin{equation}
F_{\parallel ;\bot }(\mathbf{v}_{i}) =-\frac{8\sqrt{\pi }Z^{2}e\!\!\!/^{4}n_{e}}%
{mv_{\mathrm{th}\parallel ;\perp}^{2}}\Lambda (\varkappa )\upsilon _{\parallel ;\perp
}\int_{0}^{1}\exp\left[-\frac{\upsilon_{\perp }^{2}\zeta ^{2}}{\tau} -
\frac{\upsilon_{\parallel }^{2}\zeta ^{2}}{\tau(1-\zeta ^{2})}\right] \frac{\zeta ^{2}d\zeta }{\left[\tau\left(1-\zeta ^{2}\right)\right]^{q}} ,  \label{eq:z7}
\end{equation}
where the numerical factor $q$ is the same as introduced above.

A further increase of the ion velocity at $T_{\perp }> T_{\parallel }$ finally yields
\begin{equation}
\mathbf{F}(\mathbf{v}_{i})\simeq -\frac{4\pi Z^{2}e\!\!\!/^{4}n_{e}}{mv_{i}^{2}}
\Lambda (\varkappa )\frac{\mathbf{v}_{i}}{%
v_{i}}\left[ \erf (\upsilon /\sqrt{\tau})-\frac{2}{\sqrt{%
\pi \tau }}\upsilon e^{- \upsilon^{2}/\tau}\right] \simeq -\frac{4\pi
Z^{2}e\!\!\!/^{4}n_{e}}{mv_{i}^{2}}\Lambda (\varkappa )\frac{%
\mathbf{v}_{i}}{v_{i}} ,
\label{eq:z3}
\end{equation}%
where $\upsilon^{2} =\upsilon^{2}_{\parallel} +\upsilon^{2}_{\perp} = v_{i}^2/{2} v_{\mathrm{th}\parallel}^2$ and $\erf (z)$
is the error function. At sufficiently high velocities the cooling force \eqref{eq:z3} becomes isotropic and does not depend
explicitly on the electron beam temperatures $T_{\parallel}$ and $T_{\perp}$ (see the last part of Eq.~\eqref{eq:z3}). However,
these temperatures can be involved in the generalized Coulomb logarithm in Eq.~\eqref{eq:z3}. Note that Eqs.~\eqref{eq:z1}--\eqref{eq:z3}
can be also derived from the general cooling force~\eqref{eq:x1} inserting here the distribution function~\eqref{eq:a17} and
assuming the regularized interaction potential, i.e.~$\mathcal{U} =\Lambda (\varkappa )$. Besides, Eq.~\eqref{eq:z3} completely
agrees with the asymptotic expression~\eqref{eq:x22} by taking $\mathcal{U} =\Lambda (\varkappa )$.

At $B \to 0$ and small velocities ($v_{i}<v_{\mathrm{th}\parallel ;\perp}$) the cooling forces \eqref{eq:z1}
become highly anisotropic and are given by
\begin{equation}
\mathbf{F}(\mathbf{v}_{i})\simeq -\frac{8\sqrt{\pi }Z^{2}e\!\!\!/^{4}n_{e}}{%
3mv_{\mathrm{th}\parallel }^{2}}\Lambda (\varkappa )\left[ \mathbf{n}%
\upsilon _{\parallel }\mathcal{B}_{1}(\tau )+\boldsymbol{\upsilon }_{\perp }%
\mathcal{B}_{2}(\tau )\right] ,
\label{eq:z4}
\end{equation}
where $\mathbf{n}$ is a unit vector along the axis of the electron beam anisotropy, and
\begin{eqnarray}
&&\mathcal{B}_{1}(\tau ) =\frac{3}{\tau -1}\left[ 1-\frac{1}{\sqrt{%
\left\vert 1-\tau \right\vert }}p\left( \frac{1}{\sqrt{\tau }}\right) \right] ,  \label{eq:z5} \\
&&\mathcal{B}_{2}(\tau ) =\frac{3}{2\left( \tau -1\right) }\left[ \frac{\tau
}{\sqrt{\left\vert 1-\tau \right\vert }}p\left( \frac{1}{\sqrt{\tau }}%
\right) -1\right]   \label{eq:z6}
\end{eqnarray}
with $\mathcal{B}_{1}(1)=\mathcal{B}_{2}(1)=1$, and
\begin{equation}
p(x) =\left\{
\begin{array}{cc}
\arccos x & x<1 \\
\ln (x+\sqrt{x^{2}-1}) , & x>1%
\end{array}%
\right. .
\label{eq:dr3}
\end{equation}

Now we consider the situation when the magnetic field is very strong and the electron cyclotron radius is the smallest length
scale, $\omega _{c}\lambda \gg (v_{i}, v_{\mathrm{th}\parallel ;\bot })$ and the friction force is only weakly sensitive to the
transverse electron velocities and, hence, is affected only by their longitudinal velocity spread. In this limit $\sin (\alpha t)/%
\alpha t \to 0$ and $G(t,\zeta )\to 1-\zeta^{2}$ we obtain from Eqs.~\eqref{eq:y1} and \eqref{eq:y2} after some lengthly but
straightforward calculations
\begin{equation}
F_{\parallel ;\bot }(\mathbf{v}_{i}) =-\frac{4\sqrt{\pi }Z^{2}e\!\!\!/^{4}n_{e}}{mv_{%
\mathrm{th}\parallel }^{2}}\Lambda (\varkappa )\upsilon _{\parallel ;\perp
}\int_{0}^{1}\exp \left( -\upsilon _{\parallel }^{2}\zeta ^{2}-\frac{%
\upsilon _{\perp }^{2}\zeta ^{2}}{1-\zeta ^{2}}\right) (C-2\upsilon
_{\parallel }^{2}\zeta ^{2}) \frac{\zeta ^{2}d\zeta }{(1-\zeta ^{2})^{q}} ,
\label{eq:z9}
\end{equation}
where $C=3$, $C=1$ and $q =0$, $q =1$ for $F_{\parallel} (\mathbf{v}_{i})$ and $F_{\bot} (\mathbf{v}_{i})$, respectively.
As expected the cooling forces in Eq.~\eqref{eq:z9} are independent of the transverse
temperature $T_{\bot}$ of the electrons except that $T_{\bot}$ may be involved in the Coulomb logarithm $\Lambda (\varkappa )$.

Expressions in Eq.~\eqref{eq:z9} (as well as Eq.~\eqref{eq:z1}) are very convenient for numerical calculations since they
involve one--dimensional integrals with finite range. Similar expressions have been obtained by Pestrikov \cite{pes05} where,
however, the drag force involves an integral with infinite range. But up to the definition of the Coulomb logarithm (i.e.,
$\mathcal{U} =\Lambda (\varkappa )$ in our case and $\mathcal{U} =\mathcal{U}_{\mathrm{C}}$ in Ref.~\cite{pes05}) both expressions
are identical. This can be easily shown after changing the variable $\zeta$ in \eqref{eq:z9} to $x= [\upsilon_{\parallel }^{2}\zeta %
^{2}+\upsilon _{\perp }^{2}\zeta^{2}/(1-\zeta ^{2})]^{1/2}$ and some subsequent rearrangement.

In particular, Eq.~\eqref{eq:z9} is essentially simplified for a completely flattened distribution function of the
electrons in the limit $T_{\parallel} \to 0$, i.e.~a delta--like distribution function with respect to $v_{e\parallel}$ in
Eq.~\eqref{eq:a17}. In this case it is straightforward to show that the parallel and transverse cooling forces in Eq.~\eqref{eq:z9}
are identical with Eqs.~\eqref{eq:x43} and \eqref{eq:x44}, respectively, with $\mathcal{U} =\Lambda (\varkappa)$.

In the high--velocity limit with $\omega _{c}\lambda \gg v_{i} \gg v_{\mathrm{th}\parallel ;\bot}$, the parallel and transverse
components of the cooling force, Eq.~\eqref{eq:z9}, become
\begin{eqnarray}
&&F_{\parallel }(\mathbf{v}_{i}) \simeq -\frac{\pi Z^{2}e\!\!\!/^{4}n_{e}}{%
mv_{\mathrm{th}\parallel }^{2}}\Lambda (\varkappa )\frac{\upsilon
_{\parallel }}{\upsilon ^{3}}\left\{ \frac{3\upsilon _{\perp }^{2}}{\upsilon
^{2}}\erf \left( \upsilon \right) +\frac{2}{\sqrt{\pi }}\upsilon
e^{-\upsilon ^{2}}\left[ \frac{\upsilon _{\parallel }^{2}}{\upsilon ^{2}}%
\left( 3+2\upsilon ^{2}\right) -3\right] \right\} ,  \label{eq:z11}  \\
&&F_{\bot }(\mathbf{v}_{i}) \simeq -\frac{\pi Z^{2}e\!\!\!/^{4}n_{e}}{mv_{%
\mathrm{th}\parallel }^{2}}\Lambda (\varkappa )\frac{\upsilon _{\perp }}{%
\upsilon ^{3}}\left\{ \left( 1-\frac{3\upsilon _{\parallel }^{2}}{\upsilon
^{2}}\right) \erf \left( \upsilon \right) +\frac{2}{\sqrt{\pi }}%
\upsilon e^{-\upsilon ^{2}}\left[ \frac{\upsilon _{\parallel }^{2}}{\upsilon
^{2}}\left( 3+2\upsilon ^{2}\right) -1\right] \right\} ,   \label{eq:z12}
\end{eqnarray}
With further increase of the
ion velocity we can then neglect the exponential terms in Eqs.~\eqref{eq:z11} and \eqref{eq:z12} while $\erf (\upsilon ) \to 1$
which yields the asymptotic expressions Eqs.~\eqref{eq:x43} and \eqref{eq:x44} (for $\mathcal{U} =\Lambda (\varkappa )$),
corresponding again as well to the often considered limit $T_{\parallel} \to 0$.

The forces given by Eqs.~\eqref{eq:z11} and \eqref{eq:z12} (or Eqs.~\eqref{eq:x43} and \eqref{eq:x44} with $\mathcal{U} =\Lambda
(\varkappa )$) decay as the
corresponding force \eqref{eq:z3} like $\sim v_{i}^{-2}$ with the ion velocity. But here,
the parallel force~\eqref{eq:x43} vanishes at $v_{i\perp} =0$ which is a consequence of the presence of a strong magnetic field, where the electrons move parallel to the magnetic field. If the ion
moves also parallel to the field (i.e. $v_{i\perp} =0$) the averaged friction force must vanish within the BC treatment for symmetry reasons. The sign of the transverse force~\eqref{eq:x44} depends on the angle between
ion velocity and the magnetic field and tends to defocus ions with small transverse velocity, $v_{i\perp}
<\sqrt{2}v_{i\parallel}$ while focusing them in the opposite case.

Finally we also investigate the case of small velocities at strong magnetic fields. Introducing a new integration variable
$y^{2} =\upsilon _{\perp }^{2}\zeta ^{2}/(1-\zeta ^{2})$ in Eq.~\eqref{eq:z9} and considering a small parallel velocity
($\upsilon_{\parallel }\ll 1$) we arrive at
\begin{eqnarray}
&&F_{\parallel }(\mathbf{v}_{i}) \simeq  \frac{4\sqrt{\pi }Z^{2}e\!\!\!/^{4}n_{e}}{mv_{\mathrm{th}\parallel }^{2}}%
\Lambda (\varkappa )\upsilon _{\parallel }\frac{\partial }{\partial \xi }\xi
^{2}\frac{\partial ^{2}}{\partial \xi ^{2}}\left[ e^{\xi }K_{0}\left( \xi\right) \right] \label{eq:z13} \\
&&=-\frac{2\sqrt{\pi }Z^{2}e\!\!\!/^{4}n_{e}}{mv_{\mathrm{th}\parallel }^{2}}\Lambda (\varkappa
)\upsilon _{\parallel }\upsilon _{\perp }^{2}e^{\upsilon _{\perp }^{2}/2}%
\left[ \left( 1+2\upsilon _{\perp }^{2}\right) K_{1}\left( \frac{\upsilon
_{\perp }^{2}}{2}\right) -\left( 3+2\upsilon _{\perp }^{2}\right)
K_{0}\left( \frac{\upsilon _{\perp }^{2}}{2}\right) \right] ,   \nonumber   \\
&&F_{\bot }(\mathbf{v}_{i}) \simeq -\frac{2\sqrt{\pi }Z^{2}e\!\!\!/^{4}n_{e}}{mv_{\mathrm{th}\parallel }^{2}}%
\Lambda (\varkappa )\upsilon _{\perp }\frac{\partial }{\partial \xi }\xi
\frac{\partial }{\partial \xi }\left[ e^{\xi }K_{0}\left( \xi \right) \right] \label{eq:z14}  \\
&&=-\frac{2\sqrt{\pi }Z^{2}e\!\!\!/^{4}n_{e}%
}{mv_{\mathrm{th}\parallel }^{2}}\Lambda (\varkappa )\upsilon _{\perp
}e^{\upsilon _{\perp }^{2}/2}\left[ \left( \upsilon _{\perp }^{2}+1\right)
K_{0}\left( \frac{\upsilon _{\perp }^{2}}{2}\right) -\upsilon _{\perp
}^{2}K_{1}\left( \frac{\upsilon _{\perp }^{2}}{2}\right) \right] ,  \nonumber
\end{eqnarray}
where $\xi =\upsilon _{\perp }^{2}/2$. As expected the parallel force is linear with respect to $\upsilon_{\parallel}$
decreasing with an increasing transverse component $\upsilon_{\perp}$ of the ion velocity as $F_{\parallel } \sim \upsilon^{-3}_{\perp}$.
The transverse force does not depend on $\upsilon_{\parallel }$ in this limit and falls as $F_{\perp} \sim \upsilon^{-2}_{\perp}$ with the transverse velocity.

Considering now a small transverse velocity $\upsilon_{\perp} \ll 1$ yields
\begin{eqnarray}
&&F_{\parallel }(\mathbf{v}_{i})=-\frac{4\sqrt{\pi }Z^{2}e\!\!\!/^{4}n_{e}}{%
mv_{\mathrm{th}\parallel }^{2}}\Lambda (\varkappa )\upsilon _{\parallel
}e^{-\upsilon _{\parallel }^{2}} , \label{eq:z15}  \\
&&F_{\bot }(\mathbf{v}_{i}) = -\frac{4\sqrt{\pi }Z^{2}e\!\!\!/^{4}n_{e}}{%
mv_{\mathrm{th}\parallel }^{2}}\Lambda (\varkappa )\upsilon_{\perp}
\left[e^{-\upsilon ^{2}}\left( 1-2\upsilon ^{2}\right) \ln \left(\frac{2\upsilon}{%
\upsilon_{\perp}}\right) +H(\upsilon )\right] ,    \label{eq:z16}
\end{eqnarray}%
where two functions have been introduced
\begin{eqnarray}
&&H(\upsilon ) =\frac{1}{2}e^{-\upsilon ^{2}}\left( 2\upsilon ^{2}-1\right)
\Ei (\upsilon ^{2})-1+Y^{\prime }(\upsilon ) ,  \label{eq:z17}  \\
&&Y(\upsilon ) =-2\upsilon ^{3}\int_{0}^{1}e^{-\upsilon ^{2}x^{2}}\ln \left(
1-x\right) xdx .   \label{eq:z18}
\end{eqnarray}
Here $\Ei (z)$ is the exponential integral and the prime in Eq.~\eqref{eq:z17} indicates the derivative with respect to
the argument. The function $H(\upsilon )$ at small ($\upsilon \ll 1$) and large ($\upsilon \gg 1$) values of the argument
behaves as $H(\upsilon )\simeq \ln (1/\upsilon )-1-\gamma /2$ and $H(\upsilon )\simeq -\sqrt{\pi }/2\upsilon^{3}$, respectively,
where $\gamma$ is Euler's constant. Now it is seen that at $\upsilon_{\perp} \ll 1$ the parallel force \eqref{eq:z15}
decays exponentially (i.e. much faster) with $\upsilon_{\parallel}$ in contrast to the power law decays considered above.
The transverse force \eqref{eq:z16}, on the other hand, leads at low transverse ion velocities $\upsilon_{\perp}$ to a term
which behaves as $\sim \upsilon_{\perp}\ln (1/\upsilon_{\perp})$. Thus the friction coefficient in transverse direction
diverges logarithmically at small $\upsilon_{\perp}$. This is a quite unexpected behavior compared to the well--known linear
velocity dependence without magnetic field (see asymptotic expressions above). Finally, with increasing parallel velocity
$\upsilon_{\parallel}$ of the ion the logarithmic term vanishes exponentially and the transverse force behaves as $F_{\bot}
\sim \upsilon_{\bot}/\upsilon^{3}$.

\section{Features of the cooling forces Eqs.~(\ref{eq:y1}) and (\ref{eq:y2}) and comparison with CTMC simulations}
\label{sec:s4}

In this section we study some general properties
of the cooling forces on individual ions resulting from the BC approach by evaluating Eqs.~\eqref{eq:y1} and \eqref{eq:y2} numerically.
We consider both the effects of the magnetic field and of a variation of the shape of the electron distribution on the cooling forces at various transverse velocities $v_{i\perp}$ of the ions. The density $n_{e}\simeq 10^{6}$~cm$^{-3}$
and the temperatures $T_{\parallel} \simeq 0.1$~meV and $T_{\perp} \simeq 0.11$~eV of the electron beam are the same as in
the experiments at the ESR storage ring \cite{win96,wink96,win97} (see also Sec.~\ref{sec:s5} for further details) and are
typical for many other electron cooling experiments. Thus the electron beam is strongly anisotropic with $T_{\perp} \gg
T_{\parallel}$. As an example we choose C$^{6+}$ and Xe$^{54+}$ fully stripped ions for our calculations. In all examples
considered below the regularization parameter $\lambdabar_{0} =10^{-9}$~m and thereby meets the condition $\lambdabar_{0}%
\ll b_{0}(0)$, i.e. $\lambdabar_{0}$ does not affect noticeably the cooling forces \eqref{eq:y1} and \eqref{eq:y2} at low
and medium velocities as shown in Appendix~\ref{sec:app3}.

For a BC description beyond the perturbative regime a fully numerical treatment is required. In the present cases of interest such a numerical evaluation of the cooling forces is rather intricate, but can be successfully implemented by classical trajectory Monte Carlo (CTMC)
simulations \cite{zwi99,zwi02,zwi00}. In the CTMC method the trajectories for the ion-electron relative motion are calculated
by a numerical integration of the equations of motion~\eqref{eq:a1}. The cooling force is then deduced by averaging over a
large number (typically $10^{5}-10^{6}$) of trajectories employing a Monte Carlo sampling for the related initial conditions.
For a more detailed description of the method we refer to Refs.~\cite{ner07,ner09,ner10}. Both the analytic perturbative treatment and the non-perturbative numerical CTMC simulations are based on the same BC picture and
use the same effective spherical screened interaction $U(r)$. The following comparison of these both approaches thus essentially intends to check the validity and range of applicability of the perturbative approach as it has been outlined in the preceding sections.

\begin{figure*}[tbp]
\includegraphics[width=75.0mm]{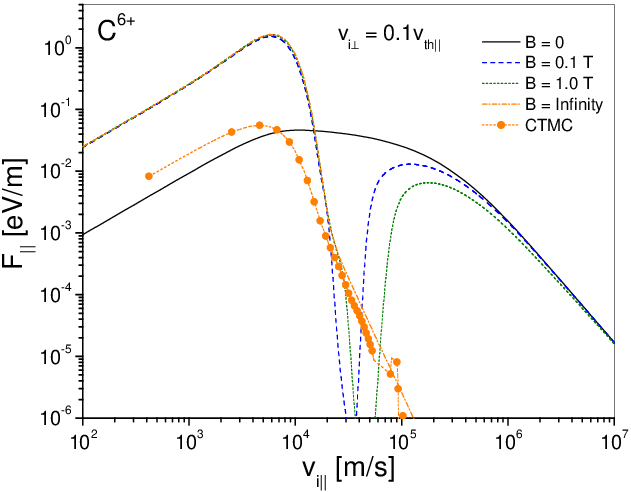}
\includegraphics[width=75.0mm]{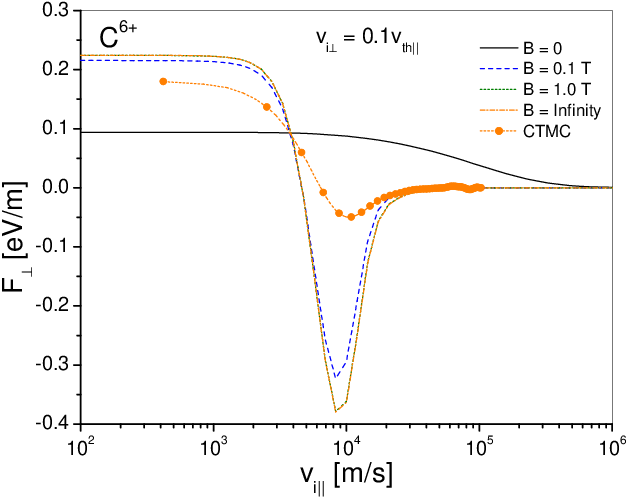}
\caption{Longitudinal ($-F_{\parallel}$, left panel) and transverse ($-F_{\perp}$, right panel) cooling forces
(with minus signs and in eV/m) for C$^{6+}$ fully stripped ions as function of the ion velocity $v_{i\parallel}$
(in m/s) and for fixed $v_{i\perp} =0.1v_{\mathrm{th}\parallel}$ and $\lambda =\lambda_{\mathrm{D}\parallel}$. The
theoretical cooling forces \eqref{eq:y1} and \eqref{eq:y2} are calculated for $\lambdabar_{0}=10^{-9}$ m (see
Appendix~\ref{sec:app2} for details) and for an electron beam with $n_{e} = 10^{6}$~cm$^{-3}$, $T_{\perp}= 0.11$~eV
and $T_{\parallel} = 0.1$~meV in a magnetic field of $B = 0$ (solid line), 0.1 T (dashed line), 1 T (dotted line),
$B = \infty$ (dash--dotted line). The CTMC results for $B = \infty$ case are shown by the filled circles. Note that
in the right panel the transverse force for $B = 0$ is increased by a factor $10^{3}$.}
\label{fig:F_mag1}
\end{figure*}

\begin{figure*}[tbp]
\includegraphics[width=75.0mm]{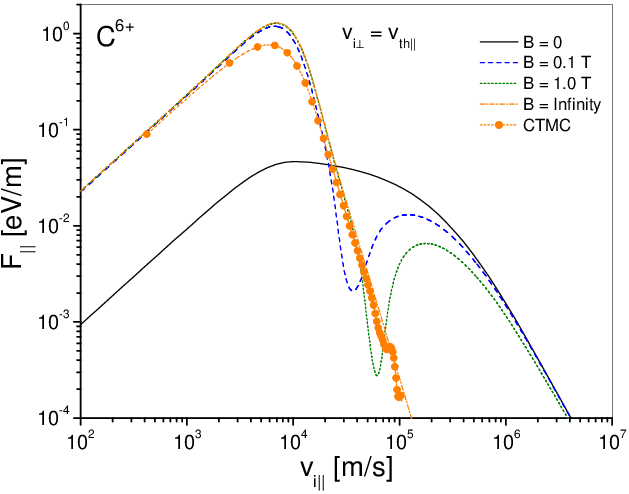}
\includegraphics[width=75.0mm]{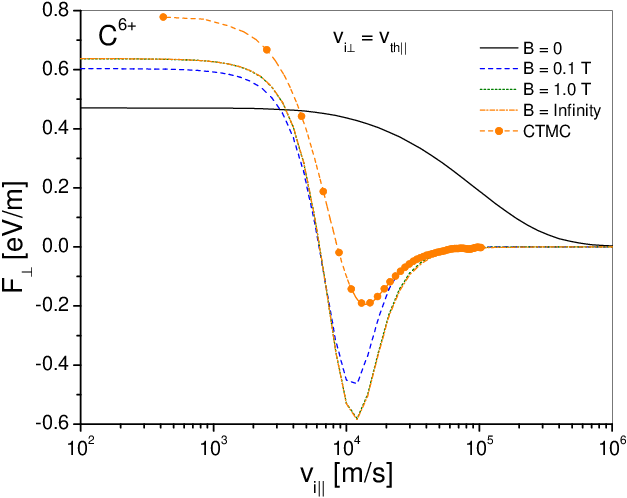}
\caption{Same as in Fig.~\ref{fig:F_mag1} but for fixed $v_{i\perp} =v_{\mathrm{th}\parallel}$.
Note that in the right panel the transverse force for $B = 0$ is increased by a factor $5\times 10^{2}$.}
\label{fig:F_mag2}
\end{figure*}

\begin{figure*}[tbp]
\includegraphics[width=75.0mm]{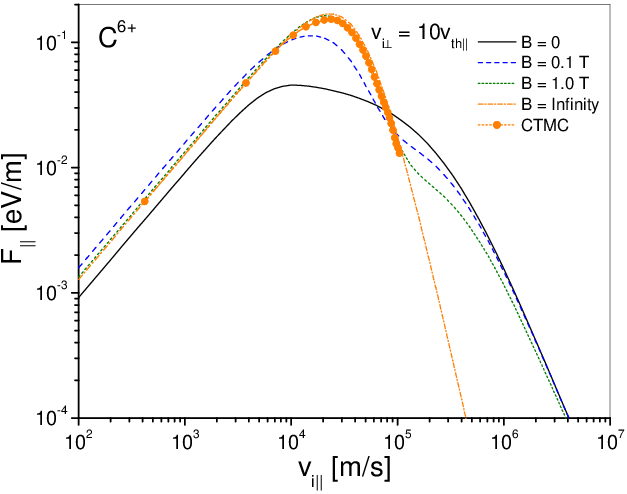}
\includegraphics[width=75.0mm]{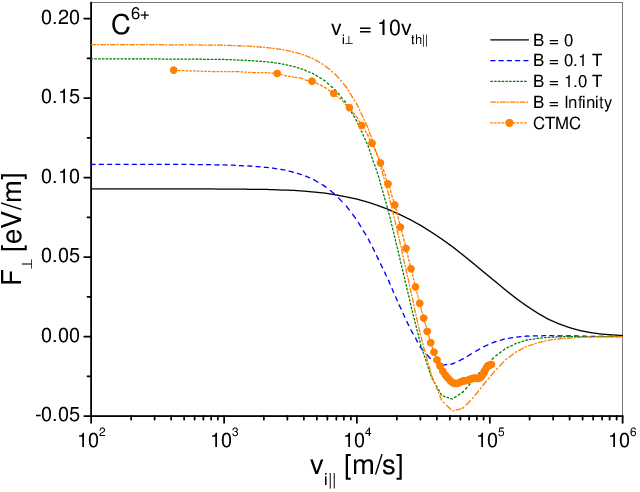}
\caption{Same as in Fig.~\ref{fig:F_mag1} but for fixed $v_{i\perp} =10v_{\mathrm{th}\parallel}$.
Note that in the right panel the transverse force for $B = 0$ is increased by a factor $10$.}
\label{fig:F_mag3}
\end{figure*}

\begin{figure*}[tbp]
\includegraphics[width=75.0mm]{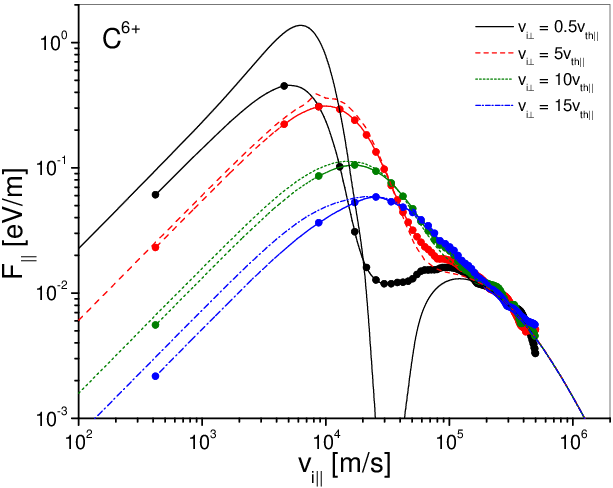}
\includegraphics[width=75.0mm]{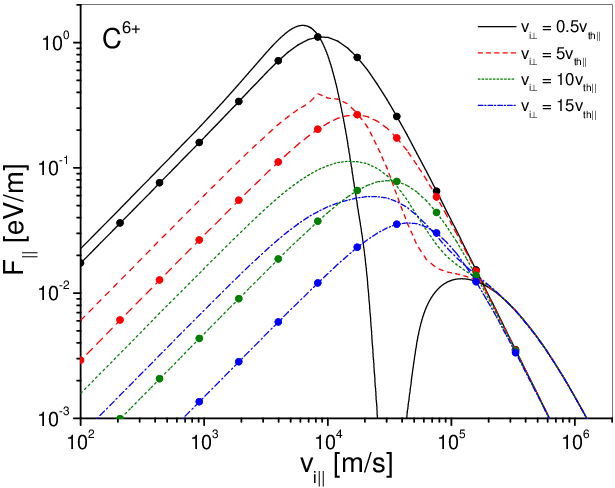}
\caption{Left panel: Longitudinal cooling force $-F_{\parallel}$ (in eV/m) for C$^{6+}$ ion as function of $v_{i\parallel}$
(in m/s) for $\lambda =\lambda_{\mathrm{D}\parallel}$ and $v_{i\perp} =0.5v_{\mathrm{th}\parallel}$ (solid lines), $v_{i\perp}=5
v_{\mathrm{th}\parallel}$ (dashed lines), $v_{i\perp} =10v_{\mathrm{th}\parallel}$ (dotted lines), and $v_{i\perp} =15
v_{\mathrm{th}\parallel}$ (dash-dotted lines). For calculation of the theoretical cooling force \eqref{eq:y1} (the lines without
symbols) the same set of parameters are used as in Fig.~\ref{fig:F_mag1} with the value of a magnetic field $B = 0.1$~T. The CTMC
results are shown by the lines with filled circles. Right panel: Same as in the left panel but the present perturbative treatment
(represented in the left panel by the lines without symbols) is compared with the PF (the lines with symbols) as given by
Eq.~\eqref{eq:force_parkhom} for $v_{\mathrm{eff}} =2v_{\mathrm{th}\parallel}$. See the text for further details.}
\label{fig:F_mag5}
\end{figure*}

\begin{figure*}[tbp]
\includegraphics[width=75.0mm]{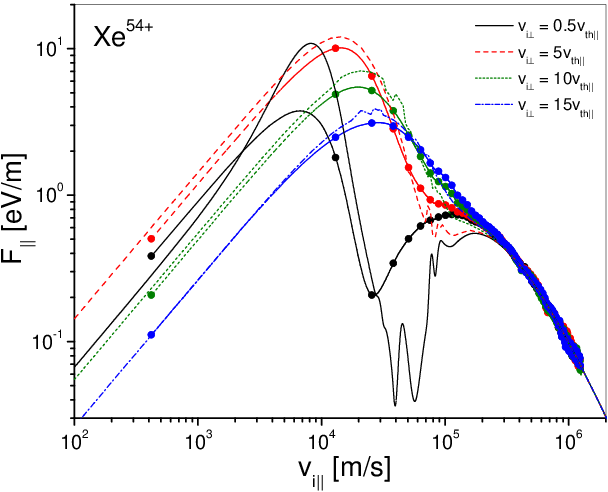}
\includegraphics[width=75.0mm]{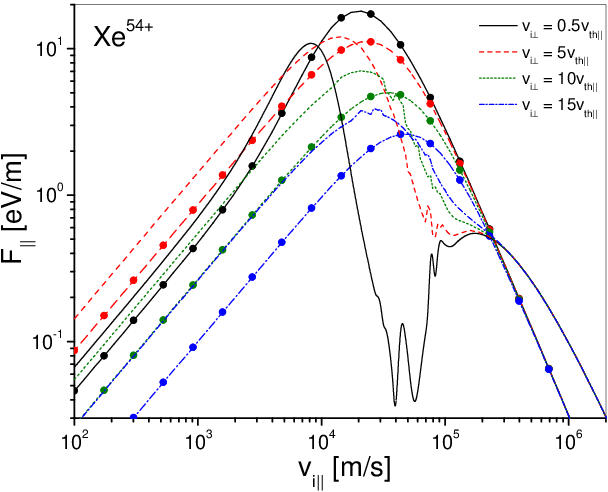}
\caption{Same as in Fig.~\ref{fig:F_mag5} but for Xe$^{54+}$ ion. In the right panel the PF is evaluated with
$v_{\mathrm{eff}} =3.5v_{\mathrm{th}\parallel}$.}
\label{fig:F_mag6}
\end{figure*}

\begin{figure*}[tbp]
\includegraphics[width=77.0mm]{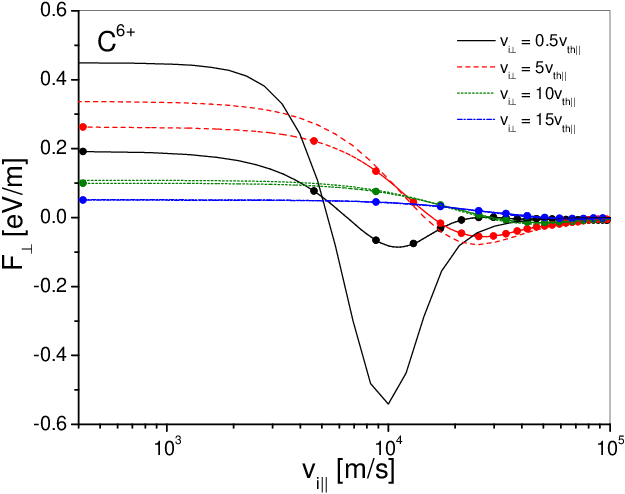}
\includegraphics[width=75.0mm]{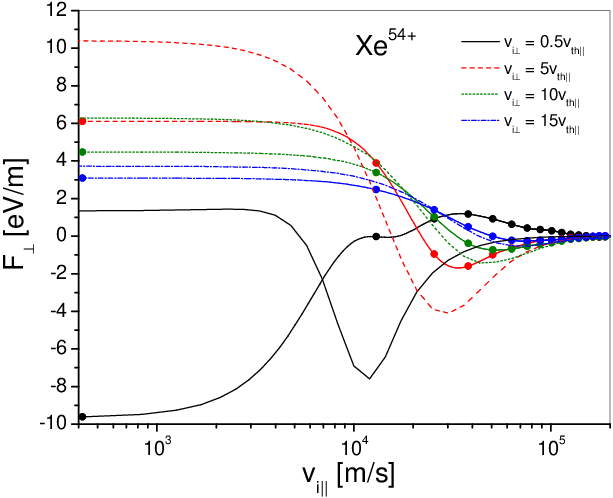}
\caption{Same as in the left panels of Figs.~\ref{fig:F_mag5} and \ref{fig:F_mag6} for C$^{6+}$ and Xe$^{54+}$
ions, respectively, but for the transverse force $-F_{\perp}$.}
\label{fig:F_mag7}
\end{figure*}

First we consider the effect of the strength of the magnetic field on the second--order cooling forces. In Figs.~\ref{fig:F_mag1}--\ref{fig:F_mag3}
the parallel ($-F_{\parallel}$, left panels) and transverse ($-F_{\perp}$, right panels) cooling forces (in eV/m) given by
Eqs.~\eqref{eq:y1} and \eqref{eq:y2}, respectively, are plotted vs ion beam parallel velocity $v_{i\parallel}$ (in m/s) for
C$^{6+}$ ions and at fixed $v_{i\perp} =0.1v_{\mathrm{th}\parallel}$
(Fig.~\ref{fig:F_mag1}), $v_{i\perp} =v_{\mathrm{th}\parallel}$ (Fig.~\ref{fig:F_mag2}) and $v_{i\perp} =10v_{\mathrm{th}\parallel}$
(Fig.~\ref{fig:F_mag3}) and for various values of the magnetic field and are shown as the lines without symbols. The two
limiting cases of vanishing ($B=0$) and infinitely strong ($B=\infty$) magnetic fields are obtained from Eqs.~\eqref{eq:z1}
and \eqref{eq:z9}, respectively. Note that the transverse
velocity of the ion is rather small, i.e.~$v_{i\perp} \ll v_{\mathrm{th}\perp}$,
in the examples Figs.~\ref{fig:F_mag1}--\ref{fig:F_mag3}, which results in a
very small transverse cooling force at $B=0$. Indeed comparing
Eqs.~\eqref{eq:z1} and \eqref{eq:z9} one concludes that typically $F_{\infty\perp}/F_{0\perp} \sim T_{\perp}/T_{\parallel} \gg 1$
at small and intermediate velocity range, where $F_{\infty\perp}$ and $F_{0\perp}$ are the transverse cooling forces at $B=\infty$
and $B=0$, respectively. Therefore in the right panels of Figs.~\ref{fig:F_mag1}--\ref{fig:F_mag3} the values of the transverse forces
at $B = 0$ are increased by some appropriate (large) factors.
The filled symbols in Figs.~\ref{fig:F_mag1}--\ref{fig:F_mag3} represent the results of the CTMC simulations obtained for
an infinitely strong magnetic field ($B=\infty$); CTMC results for a finite magnetic field are shown later in Figs.~\ref{fig:F_mag5}--\ref{fig:F_mag7}.
For simplifying the comparison, in both treatments, the perturbative BC and the CTMC calculations,
the screening length was fixed by $\lambda =\lambda_{\mathrm{D}\parallel}$,
independently of the strength of the magnetic field,
where $\lambda_{\mathrm{D}\parallel} = v_{\mathrm{th}\parallel}/\omega_{p}$ is the longitudinal Debye length and $\omega_{p}$ is the electron plasma frequency.
For the perturbative cooling forces we also employed the velocity--dependent regularization parameter $\lambdabar (v_{i\parallel})$ of the interaction potential as discussed in Appendix~\ref{sec:app2}.

Compared to the unmagnetized case with $B=0$ (solid curves in Figs.~\ref{fig:F_mag1}--\ref{fig:F_mag3}) the magnetic field
increases the cooling force $F_{\parallel}$ at low velocities while reducing it at high velocities. Furthermore, the deviations
of the parallel cooling force from the unmagnetized regime are stronger at smaller $v_{i\perp}$, that is, the cooling force is
less sensitive to $B$ at large $v_{i\perp}$ in all shown cases.

A somewhat different picture is observed for the absolute value of the transverse force, $|F_{\perp}|$ (Figs.~\ref{fig:F_mag1}--\ref{fig:F_mag3},
right panels) when turning on the magnetic field from $B=0$ to $B= \infty$. The force $F_{\perp}$ is much more sensitive to the variation of $B$ (compared to the parallel force $F_{\parallel}$) and $|F_{\perp}|$ is strongly increased by the magnetic
field in the whole parallel velocity range and for any transverse velocity $v_{i\perp}$.
While $F_{\parallel}$ is almost independent of the transverse ion velocity at small $v_{i\perp} \ll v_{\mathrm{th}\parallel}$
the transverse force $F_{\perp}$ first shows a linear increase with $v_{i\perp}$ (see Eq.~\eqref{eq:y2} and the right panels of
Figs.~\ref{fig:F_mag1} and \ref{fig:F_mag2}) but is reduced again by a further increase of $v_{i\perp}$ (Fig.~\ref{fig:F_mag3}, right panel).
In addition, for both $F_{\parallel}$ and $F_{\perp}$, a rather weak magnetic field may produce significant deviations
from the $B=0$ regime at small and intermediate velocities $v_{i\parallel}$ and $v_{i\perp}$. And at high velocities and a strong magnetic field ($B=1$ T) the cooling force $F_{\parallel}$ strongly deviates from the extreme case with $B=\infty$, which is, however, not accessible for the present experiments at storage rings.
At arbitrarily strong but finite magnetic field and sufficiently high velocities, $v_{i} \gg (\omega_{c}\lambda , v_{\mathrm{th}\parallel ;\perp})$,
the cooling force~\eqref{eq:y1} converges to the parallel unmagnetized force, Eq.~\eqref{eq:z1}, which is the leading order term $\mathcal{O} (v^{-2}_{i})$ of the high--velocity expansion of Eq.~\eqref{eq:y1}, while, as discussed in Sec.~\ref{sec:s3-2}, the regime of infinitely strong magnetic field, Eq.~\eqref{eq:z9}, is reached
for lower velocities $v_{i} \ll \omega_{c}\lambda$.
At high velocities and strong magnetic field, the cooling force given by Eq.~\eqref{eq:y1} thus deviates systematically from the regime of infinitely strong magnetic field, Eq.~\eqref{eq:z9}.

\begin{figure*}[tbp]
\includegraphics[width=75.0mm]{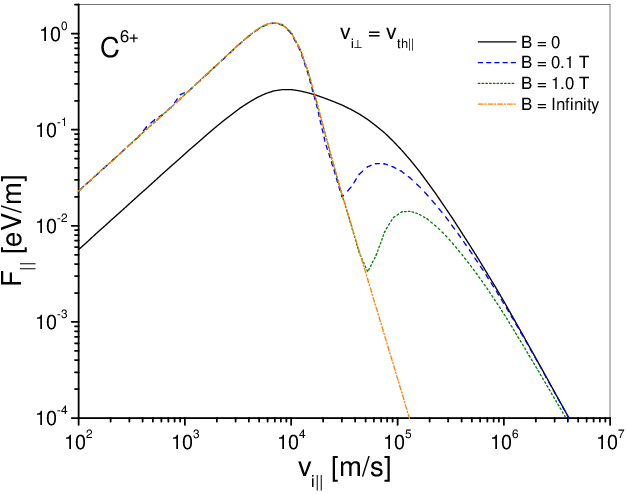}
\includegraphics[width=75.0mm]{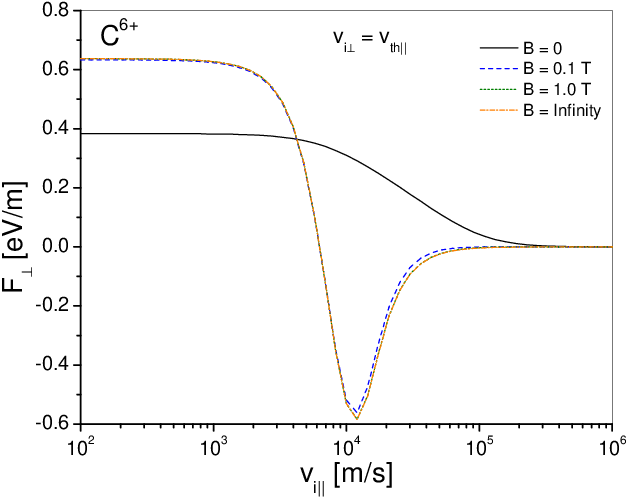}
\caption{Same as in Fig.~\ref{fig:F_mag2} with $v_{i\perp} =v_{\mathrm{th}\parallel}$ but for $T_{\perp} =10^{-2}$ eV
($T_{\perp} =10^{2}T_{\parallel}$). Note that in the right panel the transverse force for $B = 0$ is increased by a factor $20$.}
\label{fig:F_mag4}
\end{figure*}

Another interesting feature of the parallel cooling force \eqref{eq:y1} observed in Figs.~\ref{fig:F_mag1}--\ref{fig:F_mag3}, in particular at small transverse velocities $v_{i\perp}$, is the formation of two maxima at parallel ($v_{i\parallel} \sim v_{\mathrm{th}\parallel}$) and transverse ($v_{i\parallel} \sim v_{\mathrm{th}\perp}$)
electron thermal velocities with the formation of a corresponding (deep) minimum. Here, the maximum at $v_{i\parallel} \sim
v_{\mathrm{th}\parallel}$ is systematically larger than the second one at higher velocities $v_{i\parallel} \sim v_{\mathrm{th}\perp}$. And while the position of the low--velocity maximum of the force $F_{\parallel}$ is almost independent on the strength of
the magnetic field, the high--velocity maximum is reduced and its position is shifted towards higher $v_{i\parallel}$ at increasing $B$ making the force minimum deeper. A further increase of the magnetic field ($B > 1$~T) and finally the transition to the regime
$B =\infty$ results in a less structured shape of the parallel force.
Increasing, however, the transverse ion velocity $v_{i\perp}$ reduces the depth of the force minimum and results at $v_{i\perp} \gg v_{\mathrm{th}\parallel}$ in a smoother shape with only one maximum, see Fig.~\ref{fig:F_mag3} (left panel).

Figures~\ref{fig:F_mag1}--\ref{fig:F_mag3} also clearly demonstrate focusing or "antifriction" (given by the negative values shown
on the right panels of Figs.~\ref{fig:F_mag1}--\ref{fig:F_mag3} by positive values) and the change of the sign of the transverse
force $F_{\perp}$ which become more pronounced with increasing magnetic field. Similar features for the transfers force have been
reported in Refs.~\cite{fedot06,fedo06} using VORPAL simulations. The asymptotic expression~\eqref{eq:x44} (or the more accurate
asymptotic Eq.~\eqref{eq:z12}) predicts that the change of the sign of the force $F_{\perp}$ occurs at $v_{i\parallel} =v_{i\perp}%
/\sqrt{2}$ which corresponds to a constant (i.e. independent of $B$ and $v_{i\perp}$) angle $\vartheta = \arctan\sqrt 2$ between
the magnetic field $\mathbf{B}$ and the ion velocity $\mathbf{v}_{i}$. Let us recall, however, that the asymptotic expression~\eqref{eq:x44}
derived in the case of infinitely strong magnetic field is valid either at vanishing longitudinal velocity spread ($T_{\parallel} \to
0$) of the electrons or at high--velocities $v_{i\parallel} \gg v_{\mathrm{th}\parallel}$ of the ion (see Eq.~\eqref{eq:z12}). Our
numerical calculations of the second--order forces $F_{\perp}$ shown in Figs.~\ref{fig:F_mag1}--\ref{fig:F_mag3} (right panels) also
shows an almost constant angle $\vartheta$, i.e. independent of $B$, which now, however, depends on the transverse velocity $v_{i\perp}$.
At smaller $v_{i\perp}$ the angle $\vartheta$ when the force $F_{\perp}$ changes the sign is much smaller than the value predicted
by the asymptotic Eq.~\eqref{eq:x44} (Fig.~\ref{fig:F_mag1}, right panel) but with increasing $v_{i\perp}$ it converges to the constant
value given above (Fig.~\ref{fig:F_mag3}, right panel).

Comparisons of the cooling forces determined by the CTMC simulations and the second--order perturbative treatment Eq.~\eqref{eq:z9}
at infinitely strong magnetic field are presented in Figs.~\ref{fig:F_mag1}--\ref{fig:F_mag3} by the filled symbols and the dash--doted
lines, respectively. It is seen that in general the perturbative treatment overestimates the CTMC results for both components of the
cooling force which is, however, more pronounced for $F_{\perp}$. On the other hand, it is clearly observed that in the regimes of large
parallel velocity $v_{i\parallel}$ and for arbitrary $v_{i\perp}$ the second--order perturbative treatment agrees almost perfectly
(within the unavoidable numerical fluctuations) with the CTMC results. Increasing, however, the transverse velocity $v_{i\perp}$
of the ion one arrives at the regime where the conditions of the applicability of the perturbative treatment (see, e.g., the brief
discussion in Sec.~\ref{sec:s2-1}) is less critical and an excellent agreement between second--order BC and CTMC is observed in the
whole parallel velocity range as shown, for instance, in the left panels of Figs.~\ref{fig:F_mag2} and \ref{fig:F_mag3}. Obviously
the agreement between both approaches is, in general, better for the parallel forces. In addition, similar to the second--order BC
approach the CTMC also demonstrates the formation of "antifriction" for the transverse force $F_{\perp}$. That is, the second--order BC qualitatively captures the velocity domain where the force changes the sign although it does not predict
correctly the magnitude of the force at small $v_{i\perp}$.

Next we also look for some complementary information about the cooling forces \eqref{eq:y1} and \eqref{eq:y2}, and plot in
Fig.~\ref{fig:F_mag4} these forces on C$^{6+}$ ions vs ion beam parallel velocity at fixed $v_{i\perp} =v_{\mathrm{th}\parallel}$
(cf. Figs.~\ref{fig:F_mag4} and \ref{fig:F_mag2}) but for a different shape of the distribution function of the electrons with
smaller $T_{\perp} =10^{-2}$~eV. That is, Fig.~\ref{fig:F_mag4} is equivalent to Fig.~\ref{fig:F_mag2} except of the smaller transverse
thermal velocity $v_{\mathrm{th}\perp}$ and cyclotron radius $a_{\perp} =v_{\mathrm{th}\perp}/\omega_{c}$ of the electrons
in Fig.~\ref{fig:F_mag4}. This change of the transverse temperature has little influence on both components of the magnetized cooling
force, only the minimum of the parallel force is increased by decreasing $T_{\perp}$. And at this smaller cyclotron
radius $a_{\perp}$ of the electrons the transverse force is almost independent of $B$ and converges to the regime of infinitely
strong magnetic field as shown in Fig.~\ref{fig:F_mag4} (right panel). On the other hand, both components of the unmagnetized force (solid lines) are strongly increased at smaller temperature $T_{\perp}$.

The regimes of an infinitely strong magnetic field where we already compared
the CTMC simulations with the second--order perturbative treatment are, however, far from being accessible by any realistic scenario at storage rings. Thus we also present results for the second--order parallel ($-F_{\parallel}$) cooling forces (in eV/m, lines without symbols) given by Eq.~\eqref{eq:y1} as functions of the ion parallel velocity $v_{i\parallel}$ (in m/s) in Figs.~\ref{fig:F_mag5} and \ref{fig:F_mag6}, now for the fully stripped ions C$^{6+}$
and Xe$^{54+}$ at a finite magnetic field $B = 0.1$~T and fixed $v_{i\perp}=0.5
v_{\mathrm{th}\parallel}$ (solid lines), $v_{i\perp} =5v_{\mathrm{th}\parallel}$ (dashed lines), $v_{i\perp} =10v_{\mathrm{th}\parallel}$
(dotted lines), and $v_{i\perp} =15v_{\mathrm{th}\parallel}$ (dash--dotted lines).
The density and the parallel and transverse temperatures of the electron beam are the same as in the experiments at the ESR storage ring
\cite{win96,wink96,win97} (see also Fig.~\ref{fig:F_mag1}). Again, the filled symbols in the left panels of Figs.~\ref{fig:F_mag5} and
\ref{fig:F_mag6} represent the results of the CTMC simulations obtained for a magnetic field $B=0.1$~T. As before the screening length is here fixed by the constant value $\lambda =\lambda_{\mathrm{D}\parallel}$ and the velocity--dependent regularization parameter $\lambdabar (v_{i\parallel})$ needed in the perturbative BC is again as determined in Appendix~\ref{sec:app2}.

We also compared our model to the empirical formula for the parallel cooling force on a single ion
\begin{equation}
F_{\parallel}(\mathbf{v}_{i}) = - \frac{4\pi n_e Z^2 e \! \! \! /^4}{m} \,
\frac{v_{i \parallel}}{(v_{i \parallel}^2 + v_{i \perp}^2 + v_{\rm eff}^2 )^{3/2}}
\ln \left( 1+\frac{s_{\max}}{s_{\min} + a_{\perp}} \right)
\label{eq:force_parkhom}
\end{equation}
as proposed by Parkhomchuk \cite{par00,park00} (for simplicity this formula is abbreviated below as PF -- Parkhomchuk's formula).
Here $s_{\min} =|Z|e\!\!\!/^{2}/{m(v_{i}^{2}+ v_{s}^{2})}$ and $s_{\max} =(v_{i}^{2} +v_{s}^2)^{1/2}/{\omega_{p}}$
are the dynamical minimal and maximal impact parameters, respectively, $a_{\perp}$ is the cyclotron radius of the electrons,
and $v_{\mathrm{eff}}$ is an effective electron velocity related to the transverse magnetic and electric fields in the
electron cooler (see Refs.~\cite{par00,park00}) which can be viewed as a fitting parameter, and  $v_{s}$ is a characteristic thermal velocity, as discussed in Appendix~\ref{sec:app2}.
For consistency with our BC approach and the CTMC simulations, however, we
evaluated Eq.~\eqref{eq:force_parkhom} by fixing $s_{\max}$ also to the static screening length $s_{\max} =\lambda_{\mathrm{D}\parallel}$.
In the right panels of Figs.~\ref{fig:F_mag5} and \ref{fig:F_mag6}, the PF cooling forces $F_{\parallel}$
(lines with symbols, taking the rather small values $v_{\rm eff} = 2v_{\mathrm{th}\parallel}$ and $v_{\rm eff} =3.5v_{\mathrm{th}%
\parallel}$, respectively) are compared to the perturbative treatment, which is represented by the same curves (lines without
symbols) as in the left panels of Figs.~\ref{fig:F_mag5} and \ref{fig:F_mag6}.

Furthermore, in Fig.~\ref{fig:F_mag7}, we also compare second--order and CTMC results (lines without and with filled symbols,
respectively) for the transverse cooling forces $F_{\perp}$ for C$^{6+}$ (left panel) and Xe$^{54+}$ (right panel) ions for
the same set of parameters as in the left panels of Figs.~\ref{fig:F_mag5} and \ref{fig:F_mag6}.

Figures~\ref{fig:F_mag5}--\ref{fig:F_mag7} demonstrate basically the same features for the second--order cooling forces as
already discussed in connection with Figs.~\ref{fig:F_mag1}--\ref{fig:F_mag4}. Regarding the parallel components of these
forces there is a quite good overall qualitative agreement with the CTMC results.
In particular, the CTMC shows at small $v_{i\perp}$ the formation of two maxima of $F_{\parallel}$, a higher one at $v_{i\parallel} \sim v_{\mathrm{th}\parallel}$ and a lower one at $v_{i\parallel} \sim v_{\mathrm{th}\perp}$, as it is also predicted by the perturbative BC.
The perturbative BC overestimates,
however, the cooling force at low velocities as well as the depth of its minimum in between of the two maxima, with the tendency
that the quantitative agreement with the CTMC is generally strongly improved with increasing velocities $v_{i\parallel}$ and
$v_{i\perp}$ (see the left panels of Figs.~\ref{fig:F_mag5} and \ref{fig:F_mag6}).
This is basically what is to be expected for
a perturbative treatment which should work best in the high-velocity weak coupling regime as defined by Eq.~(\ref{eq:deflinregime}).
Essentially
the same behavior we also observed for the transverse force $F_{\perp}$, shown in Fig.~\ref{fig:F_mag7}, although the quantitative
agreement with CTMC is less distinct here than for $F_{\parallel}$. But again, the perturbative BC qualitatively captures well
the velocity domains where the transverse force is either negative or positive and the agreement between perturbative BC and CTMC
is clearly improved with weaker electron--ion coupling, that is, for lower $Z$ and larger $v_{i\parallel}$ and $v_{i\perp}$. The
only exception is here the case of the highly charged Xe$^{54+}$ at the lowest $v_{i\perp} = 0.5 v_{\mathrm{th}\parallel}$
(Fig.~\ref{fig:F_mag7}, right panel) where the CTMC exhibits a completely different behavior of $F_{\perp}$.
But this is also
the case of the highest electron--ion coupling parameter where, according to Eq.~(\ref{eq:deflinregime}), the applicability of a
perturbative treatment becomes questionable. In addition, for heavy ions, like e.g. Xe$^{54+}$, and low $v_{i\perp}$, that is,
for the highest electron--ion coupling, and in the vicinity of the minimum of $F_{\parallel}$ at intermediate $v_{i\parallel}$
the BC treatment starts to predict unphysical results like the sawtooth structure of $F_{\parallel}$ emerging in this domain, see the solid lines in Fig.~\ref{fig:F_mag6}.

Finally we turn to the comparisons of our model given by Eq.~\eqref{eq:y1} and the PF Eq.~\eqref{eq:force_parkhom} both shown in the
right panels of Figs.~\ref{fig:F_mag5} and \ref{fig:F_mag6}. The considerable differences between Eq.~\eqref{eq:y1} and the PF now
clearly reveal the different nature of these both approaches. The empirical PF curve shows just some shift when varying the parameters,
namely $v_{i\perp}$, while essentially retaining its shape. The perturbative BC model as well as the nonperturbative CTMC which are
based on the full equations of motion in the presence of a magnetic field exhibit a much more intricate structure, in particular at
small $v_{i\perp}$, the formation of two maxima of the parallel force $F_{\parallel}$ at parallel and transverse electron thermal
velocities. And the PF only covers the parallel force and does not offer any description of the transverse force.

\section{Cooling force for a Maxwellian ion distribution}
\label{sec:s5}

Up to now we considered the magnetized cooling force acting on the individual ion interacting with an electron beam with anisotropic velocity distribution. But often, the measured longitudinal cooling force represents an average over the drag forces on individual ions. Thus the cooling force has to be interpreted as the average $\langle F_{\parallel }(\mathbf{v}_{i})\rangle =\EuScript{F}$
of the component $F_{\parallel }(\mathbf{v}_{i})$ of the drag force
parallel to the beam axis (and
the magnetic field) over the ion distribution $f_{i}(v_{i\parallel },v_{i\bot })$ in the beam (see, e.g.,
Refs.~\cite{fedot06,fedo06,fed06,beu02}), that is,
\begin{equation}
\EuScript{F}=2\pi \int_{-\infty }^{\infty }dv_{i\parallel
}\int_{0}^{\infty }f_{i}\left( v_{i\parallel },v_{i\bot }\right)
F_{\parallel }\left( v_{i\parallel },v_{i\bot }\right) v_{i\bot} dv_{i\bot } .
\label{eq:a41}
\end{equation}%

\subsection{Averaged cooling force}
\label{sec:s5-0}

Modeling the ion beam by the anisotropic Maxwell distribution
\begin{equation}
f_{i}\left( v_{i\parallel },v_{i\bot }\right) =\frac{1}{\left(
2\pi \right) ^{3/2}\sigma _{\bot }^{2}\sigma _{\parallel
}}e^{-v_{i\bot }^{2}/2\sigma _{\bot }^{2}}e^{-\left( v_{i\parallel
}-\widetilde{v}_{i\parallel }\right) ^{2}/2\sigma _{\parallel}^{2}} ,
\label{eq:a42}
\end{equation}%
an analytic expression for the average $\EuScript{F}$, Eq.~\eqref{eq:a41}, over the BC drag force $F_{\parallel }(\mathbf{v}_{i})$
given by \eqref{eq:a18} can be derived by substituting Eqs.~\eqref{eq:a18} and \eqref{eq:a42} into Eq.~\eqref{eq:a41} and then
integrating over $v_{i\bot}$ and $v_{i\parallel}$, which yields
\begin{eqnarray}
&&\EuScript{F} (u) =-\frac{8Z^{2}e\!\!\!/^{4}n_{e} \lambda^{2}}{mv_{\mathrm{th}\parallel}^{2}}
\frac{\left( 2\pi \right) ^{4}}{4}\int_{0}^{\infty}k_{\parallel }dk_{\parallel }\int_{0}^{\infty
}U^{2} (k) k_{\bot }dk_{\bot }  \label{eq:a18-1} \\
&&\times \int_{0}^{\infty } e^ {-\frac{t^{2}}{2} \lambda^{2} \left[k_{\parallel}^{2} \delta_{\parallel}^{2}
+k_{\bot }^{2} D(t)\right]}
\left( k_{\parallel }^{2}+k_{\bot }^{2}\frac{\sin\left(\alpha t\right)}{\alpha t}\right) \sin\left(\sqrt{2}k_{\parallel}
\lambda ut\right) tdt \,.  \nonumber
\end{eqnarray}%
The introduced dimensionless parameters $D(t) =\delta^{2} +\tau \Theta (t)$,
$u=\widetilde{v}_{i\parallel }%
/\sqrt{2}v_{\mathrm{th}\parallel}$, $\delta_{\parallel}^{2}
=1+ \sigma _{\parallel }^{2}/v_{\mathrm{th}\parallel}^{2}$ and $\delta =\sigma_{\bot }/v_{\mathrm{th}\parallel}$ are related to the distribution of the ion beam \eqref{eq:a42}, where $\sigma _{\bot }^{2}=(1/2) \langle v^{2}_{i\bot} \rangle =T_{i\bot }/M$, $\sigma_{\parallel }^{2}=
\langle v^{2}_{i\parallel} \rangle - \widetilde{v}^{2}_{i\parallel} =T_{i\parallel }/M$ with the effective
transverse ($T_{i\bot }$) and longitudinal ($T_{i\parallel }$) temperatures of the ions and the ion mass $M$,
and $\widetilde{v}_{i\parallel }$ is the average cm velocity of the ion beam with respect to the electron beam.

Finally substituting the interaction potential~\eqref{eq:a19} into Eq.~\eqref{eq:a18-1} and performing the
$k_{\parallel}$--integration we arrive at
\begin{eqnarray}
&&-\EuScript{F}(u) =\frac{4\sqrt{\pi }Z^{2}e\!\!\!/^{4}n_{e}}{mv_{\mathrm{th}\parallel}^{2}}u
\int_{0}^{\infty }\frac{dt}{t}\int_{0}^{1} d\zeta \Phi (\psi (t,\zeta )) \exp \left( -
\frac{u^{2}\zeta^{2}}{P^{2}(\zeta ) }\right) \frac{1-\zeta^{2} }{P^{3}(\zeta ) Q(t,\zeta ) } \label{eq:a18a} \\
&&\times \left[ \frac{\zeta ^{2}}{P^{2}(\zeta ) }\left(3- \frac{2u^{2}\zeta^{2}}{P^{2}(\zeta )}\right)
+\frac{2\zeta ^{2}}{Q(t,\zeta )}\frac{\sin (\alpha t) }{\alpha t}\right] ,  \nonumber
\end{eqnarray}%
with $P(\zeta ) =(\delta_{\parallel}^{2}\zeta^{2}+1-\zeta^{2})^{1/2}$, $Q(t,\zeta )=D(t) \zeta^{2}+1-\zeta^{2}$. All other quantities have already been introduced in Sec.~\ref{sec:s2-3} (see above Eq.~\eqref{eq:axx}).

While Eq.~\eqref{eq:a18a} has to be evaluated numerically, closed analytic expression
can be derived for the limiting cases of \eqref{eq:a18a} at high-- and low--velocities
and strong magnetic fields. In the high--velocity limit with $\widetilde{v}_{i\parallel }>(\omega _{c}\lambda ,v_{\mathrm{th}\parallel%
;\perp} ,\sigma_{\parallel ;\perp})$ only small $t$ contribute to the cooling force \eqref{eq:a18a} and
$\sin(\alpha t)/\alpha t \to 1$ and $Q(t,\zeta )\to \delta_{\bot }^{2}\zeta^{2}+1-\zeta^{2}$, where $\delta_{\bot}^{2}=D(0)
=\delta^{2} +\tau$. At a sufficiently large ion beam velocity Eq.~\eqref{eq:a18a} then turns into
\begin{equation}
-\EuScript{F}(u)  \simeq \frac{2\pi
Z^{2}e\!\!\!/^{4}n_{e}}{mv_{\mathrm{th}\parallel }^{2}}\Lambda
(\varkappa ) \frac{1}{u^{2}}\left[ \erf%
\left( \frac{u}{\delta _{\bot }}\right) -\frac{2}{%
\sqrt{\pi }}\frac{u}{\delta _{\bot
}}e^{-u^{2}/\delta _{\bot }^{2}}\right]
\simeq \frac{2\pi Z^{2}e\!\!\!/^{4}n_{e}}{mv_{\mathrm{th}\parallel }^{2}}%
\frac{\Lambda (\varkappa )}{u^{2}} ,
\label{eq:a48}
\end{equation}
where the force decreases as $\EuScript{F}(u)\sim u^{-2}$ with the beam velocity.

At very strong magnetic fields, when the electron cyclotron radius is the smallest length scale and $\sin (\alpha t)/\alpha t \to 0$, $Q(t,\zeta )\to \delta^{2} \zeta^{2}+1-\zeta^{2}$, and in the high--velocity limit with $\omega _{c}\lambda \gg \widetilde{v}_{i\parallel}\gg (v_{\mathrm{th}\parallel
;\bot },\sigma_{\parallel ;\bot})$, we obtain
\begin{equation}
-\EuScript{F}(u) \simeq \frac{3\pi Z^{2}e\!\!\!/^{4}n_{e}}{mv_{\mathrm{th}\parallel }^{2}}\Lambda
(\varkappa ) \frac{\delta^{2}}{u^{4}}\left[\erf \left( \frac{u}{\delta}\right) -\frac{2}{3\sqrt{\pi}}
\frac{u}{\delta}\left( 3+\frac{2u^{2}}{\delta^{2}}\right) e^{-u^{2}/\delta^{2}}\right]
\simeq \frac{3\pi Z^{2}e\!\!\!/^{4}n_{e}}{mv_{\mathrm{th}\parallel }^{2}}\Lambda (\varkappa ) \frac{\delta^{2}}{u^{4}}.
\label{eq:a51}
\end{equation}%
There is an important difference if we compare Eqs.~\eqref{eq:a51} and
\eqref{eq:a48}. The force \eqref{eq:a51} decays as $\EuScript{F}(u) \sim u^{-4}$ much faster than in Eq.~\eqref{eq:a48}.
The velocity of the beam in Eq.~\eqref{eq:a51} is large but is restricted to the value $\omega _{c}\lambda $, i.e.
$1\ll u\ll \omega_{c}\lambda $. Thus it cannot be arbitrarily large. The velocity in Eq.~\eqref{eq:a48}
is arbitrarily large but now restricted below, $\widetilde{v}_{i\parallel }\gg \omega _{c}\lambda $, i.e. the magnetic
field there cannot be arbitrarily large.

Considering on the other hand also the case of small velocities $u\ll 1$ at strong magnetic fields, Eq.~\eqref{eq:a18a} becomes
\begin{eqnarray}
&&-\EuScript{F}(u) \simeq \frac{8\sqrt{\pi }Z^{2}e\!\!\!/^{4}n_{e}}{5mv_{\mathrm{th}\parallel }^{2}
\delta_{\parallel}\delta^{2}}\Lambda (\varkappa ) u \mathcal{P}\left(\frac{\delta_{\parallel }}{\delta}\right) , \label{eq:dr1} \\
&&\mathcal{P}(x) =\frac{5}{2\left( 1-x^{2}\right) ^{2}}\left[ x^{2}+2-%
\frac{3x}{\sqrt{\vert 1-x^{2}\vert }}p(x) \right] , \label{eq:dr_new}
\end{eqnarray}%
and $p(x)$ is given by Eq.~\eqref{eq:dr3}. As expected the low--velocity cooling force Eq.~\eqref{eq:dr1} strongly depends
on the details of the distribution functions of electrons and ions.

\subsection{Comparison with experiment}
\label{sec:s5-2}

With the theoretical formalism presented above, we now compare the cooling forces on the ions resulting from our analytical
approach, Eq.~\eqref{eq:a18a} with available experimental data.

\begin{figure}[tbp]
\includegraphics[width=75.0mm]{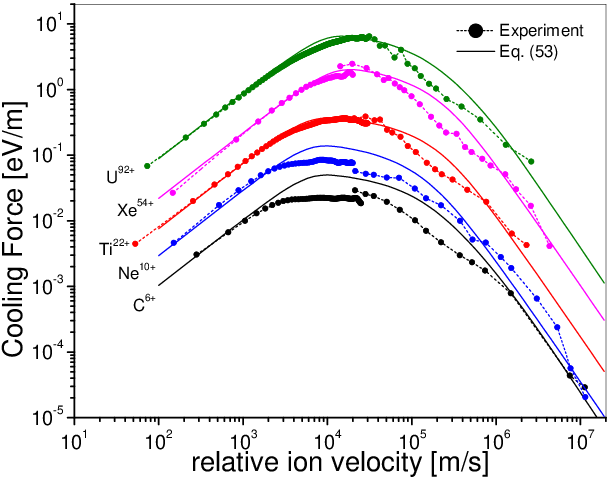}
\caption{Longitudinal cooling force (in eV/m) for various fully stripped ions as function of the
relative ion velocity (in m/s). Filled circles: experimental data from measurements at the electron cooler of the ESR
storage ring \cite{win96,wink96,win97}. Solid curves: Eq.~\eqref{eq:a18a}. The theoretical predictions of the cooling
force are calculated for an electron beam with $n_{e} = 10^{6}$~cm$^{-3}$, $T_{\perp} = 0.11$~eV and $T_{\parallel} =
0.1$~meV in a magnetic field of $B = 0.1$~T, and are fitted to the experimental results at low relative velocities by
treating the quantities $\sigma_{\parallel}, \sigma_{\perp}$ as free parameters (see the text for details).}
\label{fig:2}
\end{figure}

Measurements of the cooling forces have been performed at several storage rings, like e.g. at the ESR at GSI \cite{win96,wink96,win97}.
In these experiments a so-called cooling force is extracted, which can be viewed as a
stopping force averaged over the ion distribution in the beam and the electron distribution.
As an example we focus on the measurements of
longitudinal cooling forces for different fully stripped heavy ions as conducted at the electron cooler of the
ESR storage ring. Two different methods have been used here to determine the cooling force. At low ion velocities
the cooling force is extracted from the equilibrium between cooling and longitudinal heating with rf noise. At high
relative velocities between the rest frames of the beams the cooling force is deduced from the momentum drift of
the ion beam after a rapid change of the electron energy. Details of these methods as well as the experimental
conditions and observations are given in Refs.~\cite{win96,wink96,win97}.
The measured cooling forces are shown in Fig.~\ref{fig:2} (filled circles) for various fully stripped ions.

The electron beam in these experiments has a density of $n_e \simeq 10^6$ cm$^{-3}$ and can be
described by an anisotropic velocity distribution \eqref{eq:a17} with $T_{\perp}= m v_{\mathrm{th}%
\perp}^2 \simeq 0.11$ eV and $T_{\parallel}= mv_{\mathrm{th}\parallel}^2 \simeq 0.1$ meV as
inferred from corresponding measurements. The strength of the magnetic guiding field
was $B=0.1$ T. The measured longitudinal cooling force represents an average over the stopping forces on
individual ions. For a comparison with the theoretical model \eqref{eq:a18a} the cooling force is thus
interpreted as the average $\langle F_\parallel \rangle$ of the component $F_\parallel$ of the stopping
force \eqref{eq:a16} parallel to the beam axis (and the magnetic field) over the ion distribution
$f_{i} (v_{i\parallel},v_{i\perp})$ in the beam (see also Refs.~\cite{fedot06,fedo06,fed06,beu02}).

For low ion velocities this average is taken with respect to the transverse ion velocity only and the cooling
force depends on the parallel ion velocity, i.e.~$\langle F_{\parallel}\rangle=\langle F_{\parallel}\rangle%
(v_{i\parallel})$. In the experimental procedure used for high ion velocities the cooling force is an average
over the complete ion distribution. This average $\langle F_{\parallel}\rangle =\langle F_{\parallel}\rangle%
(\langle v_{i\parallel}\rangle)$ depends now on the velocity of the cm of the ion beam relative to the rest
frame of the electron beam $\langle v_{i\parallel}\rangle$. Both velocities are denoted as relative ion velocity
in Fig.~\ref{fig:2}. To perform the average the distribution $f_{i} (v_{i\parallel},v_{i\perp})$
must be known. However, in Refs.~\cite{win96,wink96,win97} this distribution was not determined in detail, but
there exists an estimate of the beam angular divergence $\langle\theta_{i} \rangle \lesssim 0.5$ mrad \cite{wink96}.
This yields after transformation to the rest frame of the ion beam for the transverse ion velocities $\langle %
v_{i\perp} \rangle \simeq v^{\ast}_{i\perp} \equiv\beta\gamma c\langle\theta_{i}\rangle$, where $\beta ,\gamma$
are the relativistic factors related to the beam velocity in the lab frame and $c$ is the speed of light. For the measurements
at hand with an ion energy of 250 MeV/u ($\beta =0.615$, $\gamma =1.268$) this results in $v^{\ast}_{i\perp}
\lesssim 1.17\times 10^5$ m/s.

Now we turn to the present expression for the cooling force \eqref{eq:a18a} which is shown as solid curves in Fig.~\ref{fig:2}. The velocity spread, i.e. the widths $\sigma_{\perp},
\sigma_{\parallel}$ of the ion distribution \eqref{eq:a42}, was treated as a free parameter to fit the BC stopping force to the experimental data. As the cooling force $\EuScript{F}$ is rather sensitive to a variation of $\sigma_{\perp}$ at low parallel velocities $v_{i\parallel}$ this fit is done for the linear increase of the cooling force at low relative velocities. The velocity spread of the ion beam in transverse direction used in obtaining the solid curves is $3.5v^{\ast}_{i\perp} \lesssim \sigma_{\perp} \lesssim 4.5v^{\ast}_{i\perp}$
with $\langle \theta_{i} \rangle \simeq 0.2$ mrad ($v^{\ast}_{i\perp} \simeq 4.7\times 10^4$ m/s)
which is in good agreement with the estimated beam divergence $\langle \theta_{i}\rangle$.
The spread in the longitudinal direction is here typically $\sigma_{\parallel} \lesssim 10^{-2}\sigma _{\perp}$ as it usually occurs in many
experimental situations (see, e.g., \cite{pot90,fedot06,fedo06,fed06,beu02} and references therein), in particular at the ESR storage ring \cite{win96,wink96,win97}.
In the examples considered here the regularization parameter $\lambdabar_{0}$ varies within $10^{-10}-10^{-7}$~m with $\lambdabar_{0}\ll b_{0}(0)$, i.e.
$\lambdabar_{0}$ does not affect noticeably the cooling force \eqref{eq:a18a} at low and medium velocities (see Appendix~\ref{sec:app3}).
The BC model Eq.~\eqref{eq:a18a} well agrees with the experimental cooling force at low and high velocities but somewhat overestimates the cooling force at medium velocities. These deviations are more pronounced for lower ion charge
states, but the overall behavior is essentially independent of the ion charge.

For the parameters and conditions of the considered experiments and taking into account the averages over the
electron and ion distribution functions, the domain of hard collisions and relative velocities which violate
the condition for a perturbative treatment ${|Z|e\!\!\!/^{2}}/{m v_{0}^2\lambda } < 1$ (see Eq.~\eqref{eq:deflinregime})
is rather small and thus ensures the overall applicability of our model in the present regimes. More specific,
the related characteristic velocities $({|Z|e\!\!\!/^{2}}/{m \lambda})^{1/2}$ are here $8.7\times 10^2$
m/s for $Z=6$ and $3.4\times 10^3$ m/s for $Z=92$ (taking for low ion velocities the static screening length
$\lambda = \bar{\lambda}_{\mathrm{D}}$ defined in Appendix~\ref{sec:app2}). This has to be contrasted with a
typical lower limit of the relative ion-electron velocity $v_{0}$ which is given by the parallel thermal electron
velocity $v_{\mathrm{th}\parallel} \simeq 4.2\times 10^3$~m/s when assuming low ion velocities and neglecting
the transverse component of $v_{0}$. The deviations of the perturbative BC cooling force \eqref{eq:a18a} (solid
curves) from the ESR data (filled circles) we therefore mainly ascribe to the rather unknown distribution function of the ions in the
beam which has been modeled here in the form of an anisotropic Maxwell distribution \eqref{eq:a42}. Indeed the
actual velocity spread in ion beams may essentially differ from the Maxwellian \eqref{eq:a42} and, in particular,
in some cases the recorded profiles are parabolic rather than Maxwellian \cite{fedot06,fedo06,fed06,beu02} (see
also Ref.~\cite{pot90}). For a comprehensive comparison with the measurements and a critical evaluation of
theoretical approaches a detailed knowledge of the ion distribution is indispensable.

\section{Summary}
\label{sec:sum}

In this paper we presented and discussed analytic expressions for calculating the cooling force on ions in a model of binary collisions (BC) between ions and magnetized
electrons within second-order perturbative treatment. This has been done within the framework of an improved BC theory
which involves all cyclotron harmonics of the electrons' helical motion and which is valid for any strength of the magnetic
field and in regimes where a perturbative treatment is applicable. The cooling force is explicitly calculated for a regularized
and screened Coulomb potential. Closed expressions have been derived first for monochromatic electron beams, which have been folded
with the velocity distributions of the electrons and ions. The resulting cooling force is evaluated for anisotropic Maxwell
velocity distributions of the electrons and ions. A number of limiting and asymptotic regimes of low-- and
high--velocities as well as vanishing and strong magnetic fields have been studied. The given results show that the present
model of the cooling force is very sensitive to the velocity spreads of the electrons and ions at small relative velocities.
Main limitations and uncertainties of the present BC model are: (1) the approximations concerning the electron and ion distribution functions,
(2) the use of a spherically symmetric effective interaction accounting for screening effects and hard collisions, and (3)
the underlying perturbative expansion of the equations of motion. The latter can be well justified as long as the majority of the
electron-ion collisions which contribute to the averaged final cooling force clearly meets the condition of a weak perturbation, see Eq.~\eqref{eq:deflinregime}. The use of an effective interaction, on the other hand, and the proper choice of a velocity dependent screening length clearly needs still some support from a comparison with full self-consistent simulation approaches which can treat the complete ion-target interaction in a non-perturbative way.

The here outlined BC model for the cooling force on a single ion has been compared with classical trajectory Monte Carlo (CTMC) numerical simulations and the simple empirical ansatz~\eqref{eq:force_parkhom} proposed by Parkhomchuk. It has been shown that
there is a quite good overall qualitative and in most cases also a good quantitative agreement with the CTMC results with respect to the
parallel cooling force~\eqref{eq:y1}. A similar good qualitative
agreement has been observed for the transverse force $F_{\perp}$~\eqref{eq:y2} but the quantitative agreement with CTMC is here less
distinct than for $F_{\parallel}$. In any case, however, the perturbative BC model and the nonperturbative CTMC based on the full equations of
motion in the presence of a magnetic field exhibit a much more intricate structure as
provided by the empirical ansatz~\eqref{eq:force_parkhom}.
In a further step we also compared the theoretical cooling force~\eqref{eq:a18a}, after averaging over the ion distribution function, with the experiments performed
at the ESR at GSI \cite{win96,wink96,win97}. The overall agreement of Eq.~\eqref{eq:a18a} with the experimental cooling forces is
rather good. Unfortunately a comparison of the averaged cooling force as extracted from the experiments is only little suited for
a distinct test of the accuracy of the considered model. By demonstrating the quite involved structure and character of the BC cooling
force $\mathbf{F}(\mathbf{v}_{i})$ on a single ion we showed, however, that the good agreement with the experimental data cannot
simply be considered as accidental. The remaining deviations of Eq.~\eqref{eq:a18a} from the ESR data at medium velocities, which
can be seen in Fig.~\ref{fig:2}, are therefore essentially ascribed to the deviations of the model distribution function~\eqref{eq:a42} from
the experimental distribution of the ion beam which is not known precisely.

As the main goal of this paper we suggest a more advanced analytical model for calculations of the cooling force which is appropriate
for modeling many experimental situations with moderate or strong magnetic guiding fields. The resulting cooling forces $\mathbf{F}(\mathbf{v}_{i})$ and $\EuScript{F}(u)$ can, for instance, also be tabulated in a suitable manner to be used as input for simulations of electron cooling using the BETACOOL package \cite{lav95,mad99}.
In addition, further improvement might be achieved by performing the average involved in
Eq.~\eqref{eq:a41} numerically with recorded ion beam distributions or analytically using other ion distributions
like e.g. the parabolic distribution function as it occurs in CELSIUS~\cite{fedot06,fedo06,fed06}. Systematic comparisons for different
distribution functions and other experiments on electron cooling as well as with CTMC simulations are in progress and will be reported elsewhere.

\begin{acknowledgments}
One of the authors, H.B.N., is grateful for the support of the Alexander von Humboldt Foundation,
Germany. This work was supported by the Bundesministerium f\"{u}r Bildung und Forschung (BMBF)
under contract 06ER9064.
\end{acknowledgments}

\appendix

\section{Some consequences of the Coulomb divergency}
\label{sec:app1}

As was shown by Parkhomchuk \cite{par85} in the $B\to \infty$ limit and at high--velocities one gets asymptotic expressions for
the cooling forces which essentially differ from Eqs.~\eqref{eq:x41}--\eqref{eq:x44}. Here we will briefly show that this
is a consequence of the bare Coulomb interaction and the related Coulomb logarithm $\mathcal{U}_{\mathrm{C}}$ used in previous
treatments (see, e.g., Refs.~\cite{der78,par85,fedot06}). As has been argued in Ref.~\cite{ner07} an expression similar to the
second--order force~\eqref{eq:a10} strongly depends on the order of the integrations for any singular potential, in particular for
$U=U_{\mathrm{C}}$. Such an ambiguity does not arise for any regularized potential and, for instance, Eqs.~\eqref{eq:x1} and
\eqref{eq:x4} are finite. Assuming a finite range of the potential in Eq.~\eqref{eq:a10} we have performed first an integration
with respect to the impact parameters $\mathbf{s}$ in whole two--dimensional space. Now let us derive the cooling force~\eqref{eq:x4}
first performing the $t$--integration, i.e. changing the order of the $\mathbf{s}$-- and $t$--integrations. The calculations are
straightforward. Using the trajectory corrections in the presence of an infinitely strong magnetic field derived in Ref.~\cite{ner07}
one obtains
\begin{eqnarray}
&&F_{\parallel }(\mathbf{v}_{i}) =\frac{2\pi Z^{2}e\!\!\!/^{4}n_{e}}{m}\int
\left( 2\mathcal{T}_{1}+\mathcal{T}_{2}\right) \frac{v_{i\perp
}^{2}v_{r\parallel }}{v_{r}^{5}}f(\mathbf{v}_{e})d\mathbf{v}_{e} , \label{eq:x6} \\
&&F_{\perp }(\mathbf{v}_{i}) =-\frac{2\pi Z^{2}e\!\!\!/^{4}n_{e}}{m}\int %
[(v_{i\perp }^{2}-v_{r\parallel }^{2}) \mathcal{T}%
_{1}-v_{r\parallel }^{2}\mathcal{T}_{2}] \frac{v_{i\bot }}{v_{r}^{5}}f(%
\mathbf{v}_{e})d\mathbf{v}_{e} ,  \label{eq:x7}
\end{eqnarray}
where $v_{r\parallel }=v_{e\parallel }-v_{i\parallel }$, and the functions $T_{\nu \mu }(s)$ and quantities
$\mathcal{T}_{1}$ and $\mathcal{T}_{2}$ have been introduced in Ref.~\cite{ner07},
\begin{eqnarray}
&&\mathcal{T}_{1}=\int_{0}^{\infty }T_{12}^{2}(s)sds ,  \quad
\mathcal{T}_{2}=\int_{0}^{\infty }T_{03}(s)T_{01}(s)sds , \label{eq:x8}  \\
&&T_{\nu \mu }(s)=\frac{(2\pi )^{2}}{2}\int_{0}^{\infty }U(k)J_{\nu}(ks)k^{\mu }dk . \label{eq:x9}
\end{eqnarray}

In Ref.~\cite{ner07} we have shown that $\mathcal{T}^{\mathrm{R}}_{1}=\mathcal{T}^{\mathrm{R}}_{2}=\mathcal{U}$ for any regularized
interaction potential,
where $\mathcal{U}$ is given by Eq.~\eqref{eq:x3}. Thus, inserting these values of the coefficients $\mathcal{T}^{\mathrm{R}}_{1}$
and $\mathcal{T}^{\mathrm{R}}_{2}$ into Eqs.~\eqref{eq:x6} and \eqref{eq:x7} yields Eqs.~\eqref{eq:x41} and \eqref{eq:x42},
respectively. The situation is different for any unregularized potential, as, for instance, the
Debye--like interaction potential $U(k)=U_{\mathrm{D}}(k)$ introduced in Sec.~\ref{sec:s2}. For this potential $T^{\mathrm{D}}%
_{12}(s)=(1/\lambda ) K_{1}(s/\lambda )$, $T^{\mathrm{D}}_{03}(s) =(1/s)\delta (s) -(1/\lambda^{2} ) K_{0}(s/\lambda )$, and
$T^{\mathrm{D}}_{01}(s) =K_{0}(s/\lambda )$ (see, e.g., Ref.~\cite{ner07} for details), where $K_{n}(z)$ (with $n=0,1$) are the
modified Bessel functions, and $\lambda$ is the screening length. Transition of these functions to the bare Coulomb case is
performed by taking the limit $\lambda \to\infty$. Then $T^{\mathrm{C}}_{12}(s)=1/s$ and $T^{\mathrm{C}}_{01}(s) T^{\mathrm{C}}%
_{03}(s) \to 0$ in this limit and for any nonzero value of $s>0$. Thus, in Eqs.~\eqref{eq:x6} and \eqref{eq:x7} it can be assumed
$\mathcal{T}^{\mathrm{C}}_{2} =0$ while inserting $T^{\mathrm{C}}_{12}(s)$ into Eq.~\eqref{eq:x8} and introducing the upper and
lower cutoffs yields $\mathcal{T}^{\mathrm{C}}_{1} =\mathcal{U}_{\mathrm{C}} =\ln (r_{\max}/r_{\min})$. It is easy to see that
Eqs.~\eqref{eq:x6} and \eqref{eq:x7} with $\mathcal{T}^{\mathrm{C}}_{1}$ and $\mathcal{T}^{\mathrm{C}}_{2} =0$ completely agree
with the result reported by Parkhomchuk in Ref.~\cite{par85}. However, it should be emphasized that while the integrand in
the coefficient $\mathcal{T}^{\mathrm{C}}_{2}$ tends to zero for a bare Coulomb interaction the $s$--integration of this
integrand (i.e. the coefficient $\mathcal{T}^{\mathrm{C}}_{2}$) remains singular. This is easily proved by inserting
$T^{\mathrm{D}}_{01}(s)$ and $T^{\mathrm{D}}_{03}(s)$ into Eq.~\eqref{eq:x8}. After changing the integration variable the resulting
coefficient $\mathcal{T}^{\mathrm{D}}_{2}$ is both independent of the screening length $\lambda$ and
diverges logarithmically at small $s$. Consequently, we conclude that for any unregularized potential the coefficient $\mathcal{T}_{2}$
is of the same order
 as $\mathcal{T}_{1}$ both diverging logarithmically at small $s$ (and possibly at large $s$) and the term
proportional to $\mathcal{T}_{2}$ cannot be simply neglected in Eqs.~\eqref{eq:x6} and \eqref{eq:x7} as, for instance, in
Ref.~\cite{par85}.

\section{Adjustment of the effective interaction}
\label{sec:app2}

Our results, Eqs.~\eqref{eq:y1}, \eqref{eq:y2}, and \eqref{eq:a18a}, were derived by using the screened interaction $U_{\mathrm{R}}(r)$. As already mentioned, the use and the modelling of such an effective two body interaction is a major, but indispensable  approximation for a BC treatment where the full ion-target interaction is replaced by
an accumulation of isolated ion-electron collisions. The replacement of the complicated real non-spherically symmetric potential, like the wake fields as shown and discussed in Ref.~\cite{pet91}, with a spherically symmetric one is, however, well motivated by earlier
studies on a BC treatment at vanishing magnetic field, see Refs.~\cite{zwi09,zwi02a,zwi99a}. There it was shown by comparison with 3D self-consistent PIC simulations that the drag force from the real non-symmetric potential induced by the moving ion can be well approximated by an BC treatment employing a symmetric Debye-like potential with an effective velocity dependent screening length $\lambda(v_i)$. In these studies also an recipe was given how to derive the explicit form of $\lambda(v_i)$, which turned out to be not too much different
from a dynamic screening length of the simple form $\lambda (v_{i\parallel})
=\lambda_{\mathrm{st}}[1+ (v_{i\parallel}/v_{s})^2]^{1/2}$. Here $\lambda_{\mathrm{st}} =v_{s}/\omega_{p}$ is
the statical screening length at $v_{i\parallel} =0$,
$\omega_{p}$ is the electron plasma frequency, and $v_{s}$ is a characteristic thermal velocity which depends on the temperature anisotropy of the electron beam and
the guiding magnetic field. Although no systematic studies about the use of such an effective interaction with a screening length $\lambda(v_i)$ have been made for ion stopping in a magnetized electron plasma, the replacement of the real interaction by a velocity dependent spherical one should be a reasonable approximation also
in this case. The introduced dynamical screening length $\lambda (v_{i\parallel})$ also implies the
assumption of a weak perturbation of the electrons by
the ion and linear screening where the screening length is independent of the ion charge $Ze$, which coincide with the regimes of perturbative BC, see, e.g., Ref.~\cite{zwi02a}. Therefore we do not consider here possible nonlinear screening effects. Supposing linear screening there remains the appropriate choice
of the thermal velocity $v_{s}$, which defines the static screening $\lambda_{\mathrm{st}} =v_{s}/\omega_{p}$ at low--velocities, the dynamical one $\lambda (v_{i\parallel}) =v_{i\parallel}/\omega_{p}$ at $v_{i\parallel} \gg v_{s}$ and
the velocity scale on which the transition between static and dynamic screenings takes place.

In principle the screening length $\lambda_{\mathrm{st}}$
can be calculated within the linear--response theory using the dielectric function of a temperature--anisotropic and magnetized
plasma (see, e.g., Ref.~\cite{ner00} and references therein). This approach predicts that (i) the quantity $\lambda_{\mathrm{st}}$ is, in
general, strongly anisotropic and depends on the angle $\vartheta$
between radius--vector $\mathbf{r}$ and magnetic field $\mathbf{B}$ as well as on the strength of the magnetic field and the
temperatures $T_{\parallel}$, $T_{\perp}$ of the electron plasma.
(ii) At vanishing magnetic field the screening length $\lambda_{\mathrm{st}}$ is approximately given by the longitudinal
$\lambda_{\mathrm{D}\parallel} =v_{\mathrm{th}\parallel}/\omega_{p}$ and the transverse $\lambda_{\mathrm{D}\perp} =v_{\mathrm{th}%
\perp}/\omega_{p}$ Debye lengths at $\vartheta =0$ and $\vartheta =\pi /2$, respectively (see, e.g., Ref.~\cite{ner00}). For an
average temperature $\bar{T}=\frac{1}{3}(T_{\parallel}+2T_{\bot})$ of the electrons with corresponding thermal velocity $\bar{v}_{\mathrm{th}}
=(\bar{T}/m)^{1/2}$ the static screening can be approximated by taking $\lambda_{\mathrm{st}} =\bar{\lambda} _{\mathrm{D}} =\bar{v}_{\mathrm{th}}
/\omega_{p}$, where $\bar{\lambda}_{\mathrm{D}}$ can be considered as an angular averaged screening length.
(iii) At infinitely strong magnetic field the screening length is only determined by the longitudinal temperature $T_{\parallel}$ of the electrons, $\lambda_{\mathrm{st}}= \lambda_{\mathrm{D}\parallel}$~\cite{ner00}.

The dielectric properties of a temperature--anisotropic and magnetized
plasma thus suggest to define the thermal velocity $v_{s}$ by an interpolation between $\bar{v}_{\rm th}$ at $B=0$ and $v_{\rm th\parallel}$ at $B \to \infty$, which then
covers the entire range of the variation of a guiding magnetic field, from the unmagnetized to the strongly magnetized regimes. To this end, we propose here a
 simple interpolation formula for the characteristic velocity $v_{s}$, given by
\begin{equation}
v_{s}^{2}=\frac{\bar{v}_{\mathrm{th}}^{2}+\left( \omega _{c}/\omega
_{p}\right) ^{\mu }v_{\mathrm{th}\parallel }^{2}}{1+\left( \omega
_{c}/\omega _{p}\right) ^{\mu }}
\label{eq:vs}
\end{equation}
and take $\lambda_{\mathrm{st}} =v_{s}/\omega_{p}$ as static screening length.
Here $\mu >0$ is some positive numerical factor and the strength of the magnetic field is measured by
the quantity $\omega_{c}/\omega_{p}$. From Eq.~\eqref{eq:vs} it is seen that the transition from $B=0$ to $B=\infty$ regime is faster
for larger $\mu$, where we suggest $\mu =2$ for practical applications. But the
explicit functional form of this interpolation as well as the choice of $\mu=2$ are, of course, to a certain extent discretionary. We remark, however, that Eq.~\eqref{eq:vs} is here basically given to complete our present BC treatment by providing some reasonable recipe how to determine the required parameters for modeling
the effective interaction. The results shown and discussed in Secs.~\ref{sec:s4} and \ref{sec:s5} are obtained by fixing $v_s$ to $\bar{v}_{\rm th}$ for all cases corresponding to $B = 0$ and to $v_s = v_{\rm th\parallel}$ for all examples with $B \geqslant 0.1$\,T (where $\omega_c \gg \omega_p$ for the assumed parameter regimes). These results
are therefore not affected
by the explicit form and choice of the suggested interpolation \eqref{eq:vs}.

It should be also mentioned that, depending on the specific conditions in the storage rings, the screening length $\lambda$ has to be replaced by
the radius $r_{0}$ of the electron beam if $r_{0} <\lambda$ \cite{mes94}. Also the finite time $\tau_{f}$ of flight of the beam through the cooling section may decrease the upper cutoff if $\tau_{f} < \omega_p^{-1}$ \cite{seele98}. However, the first issue is not important for our present comparisons with experimental data \cite{win96,wink96,win97}. The radius of the electron beam and the averaged screening length in these experiments are about $r_{0}\simeq 25$~mm and $\bar{\lambda}_{\mathrm{D}} \simeq 2$~mm, respectively, and thus $r_{0}\gg \bar{\lambda}_{\mathrm{D}} $ \cite{wink96}. The time $\tau_{f}$ for the ESR experimental conditions is unfortunately not significantly larger than $\omega_{p}^{-1}$ \cite{wink96}. Thus the stationary picture we use is just applicable but the finite time $\tau_{f}$ is an additional source of uncertainty for the comparison of the present theory and experiment which needs further attention.

Next we specify the parameter $\lambdabar$ which is a measure of the softening of the interaction potential at short distances.
As we discussed in the preceding sections the regularization of the potential \eqref{eq:a19} guarantees the existence of the $s$-integrations,
but there remains the problem of treating accurately hard collisions. For a perturbative treatment the change in relative
velocity of the particles must be small compared to $v_{r}$ and this condition is increasingly difficult to fulfill in the regime $v_{r}\to 0$.
This suggests to enhance the softening of the potential near the origin the smaller $v_{r}$ is.
Within the present perturbative treatment, we employ
a dynamical regularization parameter $\lambdabar (v_{i\parallel})$ \cite{ner09,ner10}, where
$\lambdabar^{2}(v_{i\parallel}) =Cb^{2}_{0}(v_{i\parallel})+\lambdabar^{2}_{0}$ and
$b_{0}(v_{i\parallel})=|Z|e\!\!\!/^{2} /m[v^{2}_{i\parallel} +\langle v^{2}_{i\bot}\rangle +v^{2}_{s}]$,
$\langle v^{2}_{i\bot}\rangle$ is the average of $v^{2}_{i\bot}$ over the ion distribution function \eqref{eq:a42}. This average
is $\langle v^{2}_{i\bot}\rangle =v^{2}_{i\bot}$ in the case of single ion considered in Secs.~\ref{sec:s2}--\ref{sec:s4}
and $\langle v^{2}_{i\bot}\rangle =2\sigma^{2}_{\bot}$ in the case of ion beam considered in Sec.~\ref{sec:s5}.
Here $b_{0}$ is the
averaged distance of closest approach of two charged particles in the absence of a magnetic field and $\lambdabar_{0}$ is some
free parameter. In addition we also introduced $C\simeq 0.292$ in $\lambdabar(v_{i\parallel})$.
In Refs.~\cite{ner09,ner10} this parameter is deduced from the comparison of the second-order scattering cross sections with an exact
asymptotic expression derived in Ref.~\cite{hah71} for the Yukawa--type (i.e., with $\lambdabar \to 0$) interaction potential. As we have
shown in Refs.~\cite{ner09,ner10} employing the dynamical parameter $\lambdabar (v_{i\parallel})$ the second-order cross sections for
electron--electron and electron--ion collisions excellently agree
with CTMC simulations at high velocities.
Also the free parameter $\lambdabar_{0}$ is chosen such that $\lambdabar_{0}\ll b_{0}(0)$, where $b_{0}(0)$ is the distance $b_{0}(v_{i\parallel})$ at $v_{i\parallel} =0$. From the definition of $\lambdabar (v_{i\parallel})$ it can be directly inferred that
$\lambdabar_{0}$ does not play any role at low--velocities while it somewhat affects the size of the cooling force at high--velocities when $b_{0}(v_{i\parallel})\lesssim \lambdabar_{0}$.
More details on the parameter $\lambdabar_{0}$ and its influence on the cooling force are
discussed in Appendix~\ref{sec:app3}.

\begin{figure*}[tbp]
\includegraphics[width=75.0mm]{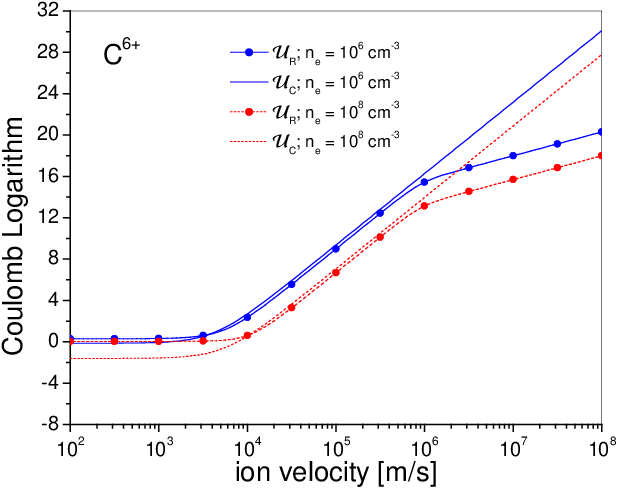}
\includegraphics[width=75.0mm]{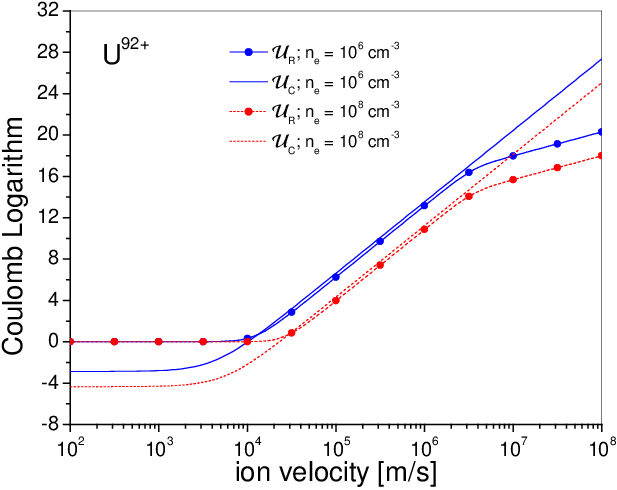}
\caption{Regularized $\Lambda (\varkappa )$ (the lines with symbols) given by Eq.~\eqref{eq:a25} and standard
$\mathcal{U}_{\mathrm{C}}$ (the lines without symbols) Coulomb logarithms for C$^{6+}$ (left panel) and U$^{92+}$
(right panel) fully stripped ions as function of $v_{i\parallel}$ (in m/s). The Coulomb logarithms are calculated
for $\lambdabar_{0} =10^{-9}$ m, $v_{i\perp} =0$, $B=0.1$~T and for $T_{\perp} = 0.11$~eV,
$T_{\parallel} =0.1$~meV, $n_{e} =10^{6}$~cm$^{-3}$ (solid lines) and $n_{e} = 10^{8}$~cm$^{-3}$ (dotted lines).}
\label{fig:log}
\end{figure*}

Our extensive numerical calculations indicated that the employed regularization parameter $\lambdabar (v_{i\parallel})$ provides a qualitatively quite
satisfactory description, although the second--order forces $F_{\parallel ;\perp}$ on a single ion are at small $v_{i\perp}$, in general, quite sensitive to variations of $\lambdabar (v_{i\parallel})$. This sensitivity is larger for highly charged ions (like, e.g., Xe$^{54+}$) and in the domain of $v_{i\parallel}$ where $F_{\parallel}$
gets its minimum (see, e.g., the deep minima in Figs.~\ref{fig:F_mag5} and \ref{fig:F_mag6}, left panels). An example of this
sensitivity is the formation of the unphysical sawtooth structure in the minimum of the parallel force shown in Fig.~\ref{fig:F_mag6}
(solid lines). Here the regularization parameter $\lambdabar (v_{i\parallel})$ is no longer capable to capture sufficiently
accurately the underlying physics.

Finally, we also illustrate in Fig.~\ref{fig:log} the features of the Coulomb logarithm
$\mathcal{U}_{\mathrm{R}}= \Lambda (\varkappa )$ given by Eq.~\eqref{eq:a25} and the standard one $\mathcal{U}_{\mathrm{C}}
= \ln (r_{\max}/r_{\min})$ for $B=0.1$~T and for different charge state $Z$ of the ions and temperatures and densities of
the electron beam close to the typical values of the experiments at the ESR storage ring \cite{win96,wink96,win97} and many
other cooling experiments. For $\mathcal{U}_{\mathrm{C}}$ we take $r_{\max} =\lambda (v_{i\parallel})$ and $r_{\min} =b_{0}
(v_{i\parallel})$. The velocity dependent lengths $\lambda (v_{i\parallel})$ and $b_{0}(v_{i\parallel})$ have been defined
and discussed above. These lengths also fix the quantity $\varkappa (v_{i\parallel})=1+ \lambda (v_{i\parallel})/\lambdabar
(v_{i\parallel})$ used for $\mathcal{U}_{\mathrm{R}}$. As can be seen from Fig.~\ref{fig:log}, at intermediate velocities
the Coulomb logarithm $\mathcal{U}_{\mathrm{R}} = \Lambda (\varkappa )$ basically shows the same behavior and features as
$\mathcal{U}_{\mathrm{C}}$, but results here in a somewhat smaller cooling force. Deviations are more pronounced at high--velocities
when the distance of the closest approach become comparable or smaller than the regularization parameter $\lambdabar_{0}$,
$b_{0}(v_{i\parallel})\lesssim \lambdabar_{0}$. It is clear that decreasing the parameter $\lambdabar_{0}$ will result in
a shift of the deviation domain shown in Fig.~\ref{fig:log} towards higher velocities. We like to emphasize, however,
that the large deviations between both Coulomb logarithms shown in Fig.~\ref{fig:log} fall in the velocity domain where the
resulting cooling forces are usually very small (see, e.g., the examples shown in Figs.~\ref{fig:F_mag1}--\ref{fig:2}).
Finally at small velocities the standard Coulomb logarithm becomes negative (i.e. $r_{\max} <r_{\min}$) which indicates
the violation of the perturbative approach, and is more pronounced either at higher densities $n_{e}$ or larger
ion charge, see Fig.~\ref{fig:log}.

\begin{figure}[t]
\includegraphics[width=75.0mm]{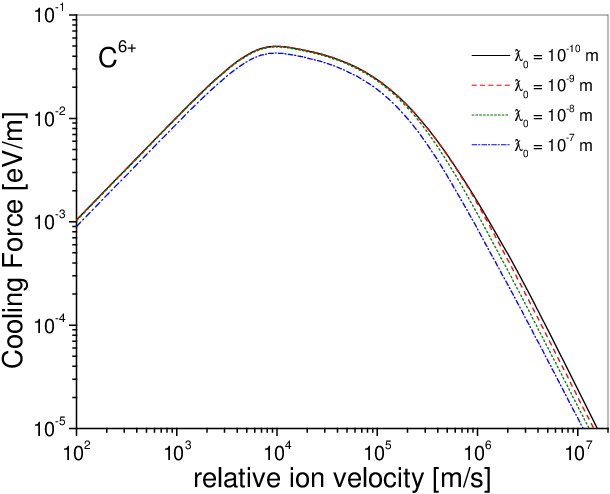}
\caption{Longitudinal cooling force (in eV/m) for C$^{6+}$ ion as function of the relative ion velocity
(in m/s). The theoretical cooling force \eqref{eq:a18a} is calculated for an electron beam with
$n_{e}= 10^{6}$~cm$^{-3}$, $T_{\perp} = 0.11$~eV and $T_{\parallel} = 0.1$~meV in a magnetic field of
$B= 0.1$ T for $\lambdabar_{0} =10^{-10}$ m (solid line), $\lambdabar_{0} =10^{-9}$ m (dashed line),
$\lambdabar_{0} =10^{-8}$ m (dotted line), $\lambdabar_{0} =10^{-7}$ m (dash--dotted line). The ion beam is
characterized by the distribution $\sigma_{\parallel} =0$, $\sigma_{\perp} =3.5 v^{\ast}_{i\perp}$,
$\langle \theta_{i}\rangle =0.2$ mrad (see Sec.~\ref{sec:s5-2} for details).}
\label{fig:lambdabar}
\end{figure}

\section{Cooling force versus the parameter $\lambdabar_{0}$}
\label{sec:app3}

Finally we briefly investigate the influence of the choice of different values of the free parameter $\lambdabar_{0}$ on
the cooling force \eqref{eq:a18a}. As mentioned in Appendix~\ref{sec:app2} this parameter is chosen such that $\lambdabar_{0}
\ll b_{0}(0)$ and therefore does not play any role at low--velocities. It adjusts, however, the cooling force in the
high--velocity regime when $b_{0}(\widetilde{v}_{i\parallel}) \lesssim \lambdabar_{0}$. Only in this high velocity limit the
parameter $\lambdabar_{0}$ directly affects (within logarithmic accuracy) the perturbative cooling force via the generalized
Coulomb logarithm $\Lambda (\varkappa )$ determined by Eq.~\eqref{eq:a25}. Thereby $\Lambda (\varkappa )$ depends on the ion
beam velocity $\widetilde{v}_{i\parallel}$ and behaves at high--velocities as $\Lambda (\varkappa ) \simeq \ln\varkappa -1 \simeq
\ln (\widetilde{v}_{i\parallel}/\omega_{p}\lambdabar_{0})-1$. This velocity dependence of $\Lambda (\varkappa )$ must be taken
into account when considering the asymptotic expressions \eqref{eq:a48} and \eqref{eq:a51}.

For the curves plotted in Fig.~\ref{fig:lambdabar} we evaluated the cooling force expression \eqref{eq:a18a} for an C$^{6+}$
ion varying the regularization parameter from $\lambdabar_{0} =10^{-10}$ m (solid line) to $\lambdabar_{0} =10^{-7}$ m (dash--dotted
line). All other parameters remain fixed and are essentially the same as in Fig.~\ref{fig:2}. For $\lambdabar_{0} \leqslant 10^{-8}$~m
the cooling force is (weakly) sensitive to a variation of $\lambdabar_{0}$, but as expected, only in the high--velocity domain.
At the larger $\lambdabar_{0} =10^{-7}$~m, where the parameter $\lambdabar_{0}$ becomes comparable to the static collision diameter,
$\lambdabar_{0} \simeq b_{0}(0)$, the cooling force shows some sensitivity to $\lambdabar_{0}$ also at low--velocities (dash--dotted
line) resulting in an overall decrease of the force. But for the higher charged ions, as considered in Sec.~\ref{sec:s5-2}, the
collision distance $b_{0}$ is larger and the sensitivity of the cooling force to $\lambdabar_{0}$ thus starts at even larger
$\lambdabar_{0}$.

\end{document}